\documentclass{aa}  

\usepackage{graphicx}
\usepackage{txfonts}
\usepackage{lipsum}
\usepackage{subcaption}         
\usepackage{lscape}            
\usepackage{placeins}
\usepackage{placeins}
\usepackage{lineno}

\begin{document}
   \title{Can dwarf novae produce type Ia supernovae?}
   \authorrunning{R. Li et al.}

   \author{R. Li\inst{1,2,3}
          ,
          C. Wu\inst{1,2}
          ,
          Z. Dai\inst{1,2,3,4}
          \and
          B. Wang\inst{1,2}
          }

   \institute{Yunnan Observatories, Chinese Academy of Sciences (CAS), Kunming 650216, China\\
              \email{[wuchengyuan; wangbo]@ynao.ac.cn}
        \and
            International Centre of Supernovae, Yunnan Key Laboratory, Kunming 650216, China
        \and
            University of Chinese Academy of Sciences, Beijing 100049, China
        \and
            Center for Astronomical Mega-Science, Chinese Academy of Sciences, Beijing 100012, China
             }
             
  \abstract
   {Accreting white dwarfs (WDs) are considered among the most promising progenitor candidates of type Ia supernovae (SNe Ia). Dwarf novae (DNe) are a subclass of cataclysmic variables (CVs) consisting of a CO WD accretor and a low-mass donor star, which can be either a main-sequence (MS) star or a slightly evolved subgiant. Previous studies suggested that, under the thermal–viscous disk instability mechanism, the time-averaged accretion rate in long-period DNe may approach the regime of stable hydrogen burning, potentially allowing WD mass growth toward the Chandrasekhar mass (\(M_{\rm Ch}\)). However, whether such periodic accretion can sustain stable H burning and lead to a mass increase of the WD remains uncertain.}   
   {We aim to explore whether high accretion rates on short, periodic timescales can maintain stable H burning on the WD surface and drive its growth toward \(M_{\rm Ch}\).}
   {We investigate the mass retention efficiency of WDs undergoing intermittent, DN-like accretion using MESA simulations, and explore the dependence on duty cycle, accretion rate, and initial WD properties.}
   {We find that periodic accretion fails to maintain stable H burning. During quiescent phases, the WD cools and becomes increasingly degenerate, triggering nova outbursts with declining mass retention efficiency, ultimately preventing further WD mass growth. We therefore suggest that DNe are unlikely to be progenitors of SNe Ia.}
   {}

   \keywords{Dwarf novae -- Type Ia supernovae -- Stellar evolution --Accretion disk instability}

   \maketitle\nolinenumbers

\section{Introduction} \label{sec:intro}
 Type Ia supernovae (SNe Ia), with typical bolometric luminosities on the order of $10^{43}$ erg/s, are among the most energetic in the universe. As reliable standard candles, due to their high luminosity and the fact that their light curves can be standardized to yield a uniform peak luminosity, SNe Ia enabled the discovery of the accelerating expansion of the universe driven by dark energy \citep{riess1998,perlmutter1999}. They are also major contributors to iron enrichment in galaxies, producing approximately 2/3 of the Milky Way's iron content \citep{greggio1983,matteucci1986,dwek2016iron}. By injecting kinetic energy and turbulence into the interstellar medium, SNe Ia drive large-scale galactic winds, stimulate star formation, and influence galaxy evolution, helping understand galactic dynamics and evolution (e.g. \citealt{joung2009dependence,strickland2009supernova,osaki1974accretion}). Additionally, as remnants of binary stellar evolution, they provide valuable insights into mass transfer, accretion processes, and binary system dynamics. Despite their importance, no pre-explosion counterparts have been detected to date, leaving the progenitors of SNe Ia uncertain \citep{hillebrandt2000type, roepke2005full, podsiadlowski2008nuclear, wang2012progenitors, maoz2012type, maoz2017star}.

 Two competing progenitor models dominate current debates: the single degenerate (SD) model and the double degenerate (DD) model (e.g., \citealt{maoz2014observational, wang2018mass, liu2023type,Ruiter2025A&ARv..33....1R}). The SD model postulates that a white dwarf (WD) accretes material from a non-degenerate companion, which can be either a main-sequence star or slightly evolved subgiant (the WD+MS channel), a helium star (the WD+He channel), or a red giant (the WD+RG channel) (see, e.g.  \citealt{Langer2000A&A...362.1046L,Han2004MNRAS.350.1301H,wang2015super,brooks2016carbon,wu2019off,wong2019evolution,wu2020formation}). Stable burning of accreted material gradually increases the WD's mass toward  the Chandrasekhar mass limit (\(M_{\rm Ch}\approx 1.4 \, \rm M_\odot\) ), eventually producing an SN Ia \citep{hoyle1960nucleosynthesis, whelan1973binaries,nomoto1984accreting,branch1985accreting}. The WD explosion with \(M_{\rm Ch}\) can reproduce the observed light curves and spectroscopy of most SNe Ia \citep{leung2018explosive}. The original DD model posits that two WDs merge via gravitational wave radiation, losing orbital angular momentum. If the resulting merged WD exceeds \(M_{\rm Ch}\), a thermonuclear explosion is triggered, producing an SN Ia  \citep{iben1984supernovae, webbink1984double}. However, some studies suggested that such mergers may instead lead to accretion-induced collapse (AIC), resulting in the formation of a neutron star (e.g. \citealt{saio1985evolution, saio1998inward, timmes1994conductive, mochkovitch1997merging,schwab2016evolution,wu2018accreting,wang2018single,wu2019off,liu2020formation,schwab2021evolutionary}), while others showed that, when rotation was taken into account, CO WD mergers could still produce SNe Ia (e.g. \citealt{yoon2007}). More recent work indicates that SNe Ia in this channel can arise from sub-\(M_{\rm Ch}\) configurations. In the violent merger scenario, two comparable mass WDs undergo a dynamical merger that directly triggers carbon detonation (\citealt{Pakmor2010,pakmor2011, pakmor2012}), and the double detonation scenario, where a helium shell detonation initiates a secondary carbon detonation in the core (\citealt{ fink2007, fink2010, boos2021}). However, the ignition conditions for violent merger and the required helium shell properties for double detonation remain uncertain.
 
In the SD channel, a cataclysmic variable (CV) system consists of a carbon–oxygen white dwarf (CO WD) accretor and a donor star that fills its Roche lobe and transfers material to the WD. The donor is typically a low-mass main-sequence (MS) star or a slightly evolved subgiant. Dwarf novae (DNe) are a subclass of CVs characterized by alternating long quiescent phases and short outbursts \citep{osaki1974accretion}. According to the thermal-viscous disk instability mechanism (\citealp{meyer1981elusive, king1997disc}), the accretion disk remains cool and dominated by neutral H, suppressing angular momentum transport. This low-viscosity state causes material to accumulate in the outer disk, corresponding to the quiescent phase. As density increases, gravitational compression heats the disk, ionizing H and enhancing viscosity. The increased viscosity enables the accumulated material in the outer disk to be efficiently transported inward, eventually accreting onto the WD surface. The rapid accretion then releases gravitational energy, producing an outburst. As the accreted material is depleted, the disk cools and returns to quiescence. This allows mass to build up again in the outer disk, setting the stage for the next instability and thus establishing the recurrent cycle between outburst and quiescence.
 
 Based on the disk instability mechanism, \citet{king2003new} proposed an evolutionary pathway for the forming of SNe Ia, in which long-period DNe (\(\gtrsim2\)d) could enable a WD to grow in mass and reach \(M_{\rm Ch}\). Although the mass transfer rate \({\dot{M}_2}\) from the donor in DNe is typically too low to sustain stable H burning, accretion onto the WD occurs only during the outburst phase. Considering the duty cycle \(d\) (defined as the ratio of the burst duration \(\tau_{\rm bur}\) to the quiescent phase duration \(\tau_{\rm qui}\), i.e. \(d = \tau_{\rm bur}/\tau_{\rm qui}\)), the equivalent accretion rate onto the WD (\(\dot{M}_{\rm acc} \sim {\dot{M}_2/d}\)) can reach the stable H burning regime. Consequently, the WD can stably burn H and retain a substantial fraction of the transferred mass during DNe outbursts, allowing it to grow in mass and potentially reach \(M_{\rm Ch}\). \citet{xu2009evolution} adopted a similar idea, suggesting that in DNe systems with \(d = 0.003\), a WD undergoing AIC could explain the pulsar GRO J1744-28. \citet{wang2010progenitors} incorporated the disk instability mechanism into their studies to estimate the birthrates and delay-time distributions of SNe Ia (see also \citealt{meng2010comprehensive}). 

 However, whether the cumulative effect of multiple short-timescale accretions can sustain long-term stable H burning remains poorly understood. \citet{nomoto2007thermal} suggested that a H envelope mass of approximately \(10^{-5} \, \rm M_\odot\) is required to sustain stable H burning on a \(1.0 \, \rm M_\odot\) WD. The timescales of DNe outbursts are extremely short, ranging from tens of days to several decades. Therefore, even at high accretion rates, the material accumulated during quiescence may be insufficient to reach the critical mass required for stable H ignition. Furthermore, the WD may cool and become more degenerate during the long quiescent period. In this case, once ignition occurs, a thermonuclear runaway may be triggered, leading to a nova explosion rather than steady burning.

 The aim of this study is to investigate whether high accretion rates over periodic short timescales can ignite the accreted material on the WD surface, allowing stable nuclear burning and enabling the WD to grow to \(M_{\rm Ch}\), thus leading to the formation of an SN Ia. In Section \ref{sec:methods}, we describe the methods used in this study. The numerical results are presented in Section \ref{sec:results}. A brief discussions and conclusions are provided in Section \ref{sec:discussion}.
 
 \section{Methods} \label{sec:methods}
 \subsection{Mass transfer and accretion-disc instability}
\begin{figure}[htbp]
    \includegraphics[width=0.45\textwidth]{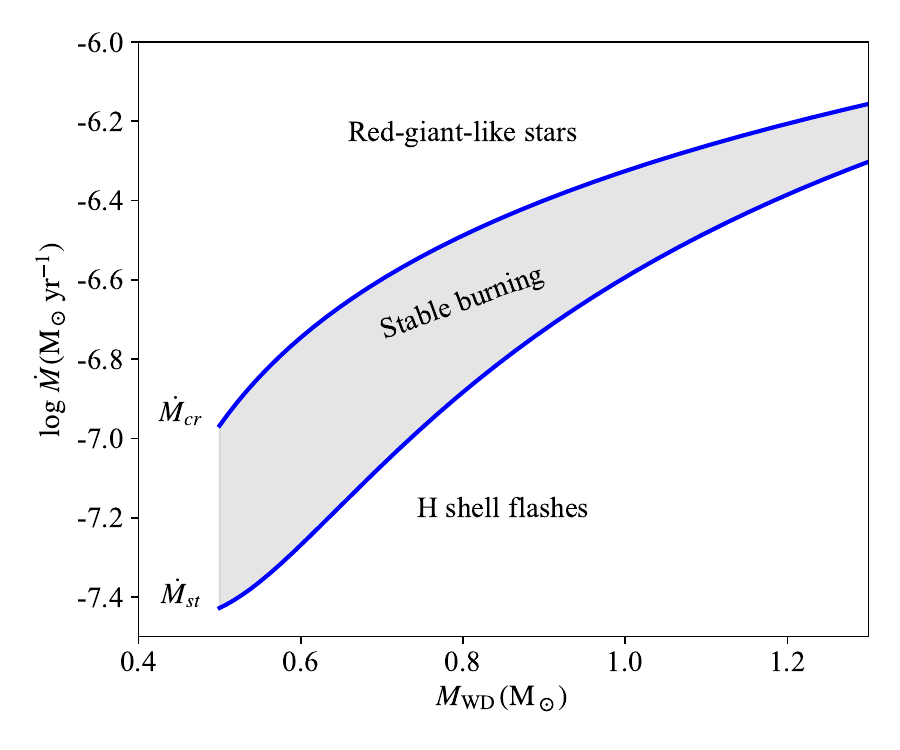} 
    \centering
    \caption{ The stable H-rich burning regime in the $\dot{M}$–$M_{\rm WD}$ plane. The horizontal axis represents the WD mass and the vertical axis the accretion rate. The upper and lower boundaries are given by Eqs.~(\ref{eq:M_cr}) and (\ref{eq:M_st}), following \cite{wang2018mass}.}
    \label{fig:mdot_in} 
\end{figure}

  The stable burning of accreted H occurs only within a specific range of accretion rates. According to \citet{wang2018mass}, the upper and lower limit of stable H burning (\(\dot{M}_{\rm cr}\), \(\dot{M}_{\rm st}\)) are as follows: 
\begin{eqnarray}
\frac{\dot{M}_{\mathrm{cr}}}{\mathrm{M_\odot\,yr^{-1}}}
= 0.27 \times10^{-7}
\left[\left(\frac{M_{\mathrm{WD}}}{\mathrm{M_\odot}}\right)^{2}
+25.52 \left(\frac{M_{\mathrm{WD}}}{\mathrm{M_\odot}}\right)-9.02\right],
\label{eq:M_cr}
\end{eqnarray}

\begin{eqnarray}
\frac{\dot{M}_{\mathrm{st}}}{\mathrm{M_\odot\,yr^{-1}}}
&=& 2.93 \times 10^{-7}
\Bigg[
-\left(\frac{M_{\mathrm{WD}}}{\mathrm{M_\odot}}\right)^{3}
+4.41 \left(\frac{M_{\mathrm{WD}}}{\mathrm{M_\odot}}\right)^{2}
\nonumber \\
&& -3.38 \left(\frac{M_{\mathrm{WD}}}{\mathrm{M_\odot}}\right)
+0.84
\Bigg],
\label{eq:M_st}
\end{eqnarray}
According to Eqs.~\ref{eq:M_cr} and \ref{eq:M_st}, Fig.~\ref{fig:mdot_in} shows the variation of the stable H-rich burning interval with the WD mass. When the accretion rate of H-rich material onto the WD ($\dot{M}_{\rm acc}$) exceeds the critical rate $\dot{M}_{\rm cr}$, the continuous accumulation of accreted material leads to expansion of the outer envelope, and the WD may enter a red-giant-like phase. In contrast, if $\dot{M}_{\rm acc}$ falls below the lower stability limit $\dot{M}_{\rm st}$, H burning becomes thermally unstable, giving rise to repeated H shell flashes that manifest as nova eruptions.

Considering that nova eruptions may occur, we include a super-Eddington stellar wind during these phases (see \citealp{paxton2010modules, paxton2015modules} for more details). The Eddington luminosity (\( L_{\text{Edd}} \)) is expressed as:
\begin{eqnarray}
L_{\text{Edd}} = \frac{4 \pi \mathrm{G} \mathrm{c} M}{\rm \kappa},
\end{eqnarray}
where \(\rm G \) is the gravitational constant, \( M \) is the mass of the WD, \(\rm c \) is the speed of light, and \(\rm \kappa \) is the Rosseland mean opacity on the surface of the WD.  

When the radiative pressure at the surface of the WD becomes sufficiently high (i.e. \( L > L_{\text{Edd}} \)), material may be ejected from the WD surface in the form of strong stellar winds. The mass-loss rate of the stellar wind (\( \dot{M}_{\text{wind}} \)) can be estimated using the empirical relation:
\begin{eqnarray}
\dot{M}_\mathrm{wind}  =  \frac{2 \eta_{\mathrm{Edd}}\left(L_{\mathrm{eff}}-L_{\mathrm{Edd}}\right)}{v_{\mathrm{esc}}^{2}},
\end{eqnarray}
where \(\eta_{\text{Edd}}\) represents the super-Eddington factor, and \(v_{\text{esc}} = \sqrt{\frac{2GM_{\text{WD}}}{R_{\text{WD}}}}\) denotes the escape velocity, \(R\) is the surface radius. The parameter \(\eta_{\text{Edd}}\) regulates the efficiency of mass loss due to the wind and takes values between 0 and 1. In our simulations, when the super-Eddington wind is activated, we assume that all energy exceeding the Eddington luminosity is utilized for mass ejection, corresponding to \(\eta_{\text{Edd}} = 1\).

The disk instability mechanism is the primary theoretical framework for explaining DN outbursts. When the mass transfer rate \({\dot{M}_2}\) from the donor star to the WD falls below a critical value \(\dot{M}_{\rm d,cr}\), the outer edge of the accretion disk enters a cool, low-viscosity state with an effective temperature below the H ionization threshold (\(\sim 6500 \, \text{K}\)). During the quiescent phase, matter gradually accumulates in the accretion disk, causing the surface density to steadily increase. As the accumulated mass grows, the disk temperature approaches the H ionization threshold. The onset of partial ionization sharply increases the local opacity, suppressing radiative cooling and disrupting the local thermal equilibrium. This triggers a thermal-viscous instability, causing the disk to transition from a cool, low-viscosity state to a hot, high-viscosity state. The material stored in the outer disk is then rapidly transported inward and accreted onto the WD surface, giving rise to a dwarf nova (DN) outburst. The critical mass transfer rate (\(\dot{M}_{\rm d,cr}\)) is given by \citep{van1996accretion}:
\begin{eqnarray}
\dot{M}_{\rm d,cr} = 4.3 \times 10^{-9} \left(\frac{P_{\text{orb}}}{4 \, \text{hr}}\right)^{1.7} \rm M_\odot \, \text{yr}^{-1},
\label{eq:disk}
\end{eqnarray}
where \( P_{\text{orb}} \) represents the orbital period of the system.

\begin{figure}[htbp]
    \includegraphics[width=0.45\textwidth]{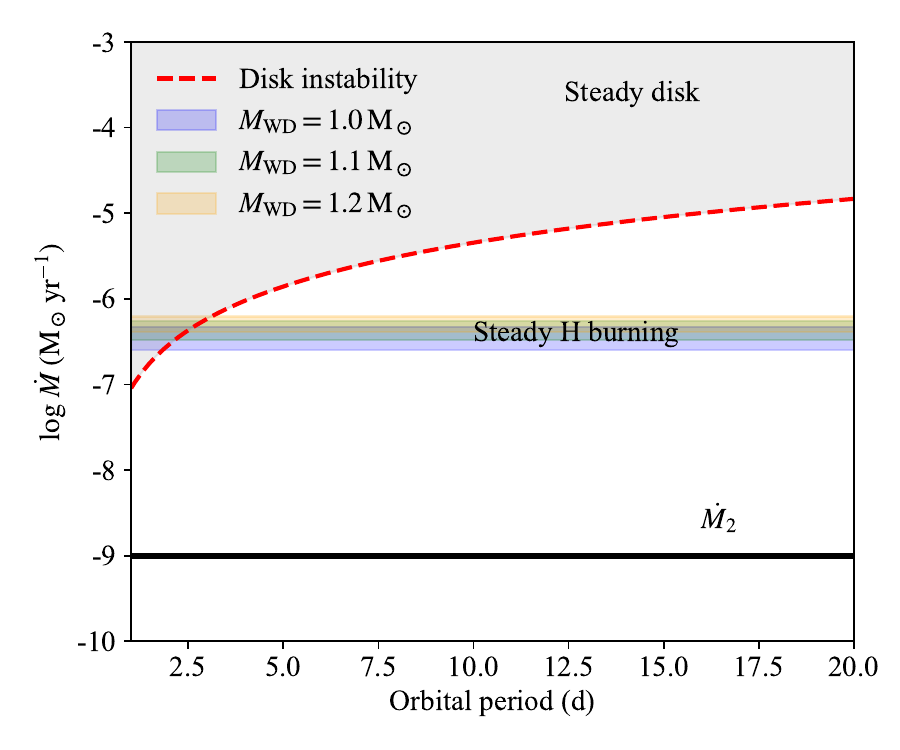} 
    \centering
    \caption{The red dashed line represents the disk instability boundary calculated from Eq.\ref{eq:disk}, with the region above this line corresponding to steady disks. The blue, red, and green rectangular regions indicate the stable H-rich burning zones of WDs with \(M_\mathrm{WD} = 1.0, 1.1, 1.2\, \rm M_\odot\). The horizontal bar labeled \(\dot{M}_2\) denotes the typical mass transfer rates observed in dwarf novae.} 
    \label{fig:mdot} 
\end{figure}

If the mass transfer rate \(\dot{M}_2\) is lower than \( \dot{M}_{\rm d,cr} \), the accretion disc becomes unstable and undergoes periodic transitions between outburst and quiescent states. During the outburst phase, the accumulated material in the disk is rapidly transferred onto the surface of the WD, while during the quiescent phase, matter accumulates in the accretion disk. Due to this instability, the WD accretes material only during outburst phases at a rate of \( \dot{M}_{\text{acc}} \), in such case \(\dot{M}_{\text{acc}} = \frac{\dot{M}_2}{d},\,\text{if } \dot{M}_2 \leq \dot{M}_{d,\text{cr}}\), where \( d \) is the duty cycle, defined as ratio of the burst duration to the quiescent phase duration. \(\dot{M}_2\) in long-period DNe is typically low (\(\sim 10^{-9} \, \rm M_\odot \, \text{yr}^{-1} \)), while \( d \) generally satisfies \( d \ll 1 \), the burst timescale \(\tau_{\rm bur}\) is on the order of days, whereas the quiescent timescale \(\tau_{\rm qui}\) is on the order of years. Under these conditions, the equivalent accretion rate \(\dot{M}_{\text{acc}}\) can reach values within the steady H-burning regime (see Fig.~\ref{fig:mdot}).

\subsection{stellar evolution code}

We employed Modules for Experiments in Stellar Astrophysics (MESA; see \citealp{paxton2010modules,paxton2013modules,paxton2015modules,paxton2017modules,paxton2019modules}) [version, r24.03.01] to simulate the accretion of H-rich material onto a CO WD periodically. For the chemical composition of the accreted material, we adopted solar abundances (\( X = 0.70 \), \( Y = 0.28 \), \( Z = 0.02 \)). We employed the Type 2 opacity table from the OPAL opacity controls to account for the chemical composition variations within the stellar interior \citep{iglesias1996updated}.

We first constructed the 1 \(\rm M_\odot\) CO WD using make\_co\_wd, starting from the \(5.9\,\rm M_\odot\) zero-age main sequence (ZAMS). The star evolved through the main sequence and post-main-sequence phases until a C/O core formed at the center. During the asymptotic giant branch (AGB) phase, the H envelope was removed manually, leaving behind a cool CO WD with a luminosity of \( 1\, \rm L_{\odot} \) and an extremely thin H envelope (\(\sim 10^{-6} \,\rm M_{\odot}\)). 

In order to stabilize the H-rich envelope, this CO WD initially accreted at a high and stable accretion rate of \( \dot{M}_{\text{acc}} = 2.5 \times 10^{-7} \, \rm M_\odot \, \text{yr}^{-1} \) for 2,000 years. It then alternated between a quiescent state (\(\tau_{\text{qui}} = 10,000\) days) with no mass accretion and an outburst state (\(\tau_{\text{out}} = 100\) days) characterized by an equivalent accretion rate of \( \dot{M}_{\text{acc}} = 4 \times 10^{-7} \, \rm M_\odot \, \text{yr}^{-1} \), corresponding to \(\dot{M}_2 = 4 \times 10^{-9} \, \rm M_\odot \, \text{yr}^{-1} \), yielding a duty cycle of \( d = 0.01 \). This accretion rate lies within the stable H-burning regime defined in Eqs.~\ref{eq:M_cr} and \ref{eq:M_st} and corresponds to a relatively high value within that range. The DN-like accretion is prescribed externally and maintained throughout the entire evolution, including during nova outbursts. Since the timescale of nova outbursts is much shorter than the quiescent phase of the DN cycle, the outburst phase is not affected by the imposed accretion prescription. \footnote{The complete list of controls is available to the reader as our MESA input files are posted online at \url{https://doi.org/10.5281/zenodo.19585622}.}

\section{Results} \label{sec:results}

\begin{figure}[htbp]
    \includegraphics[width=0.45\textwidth]{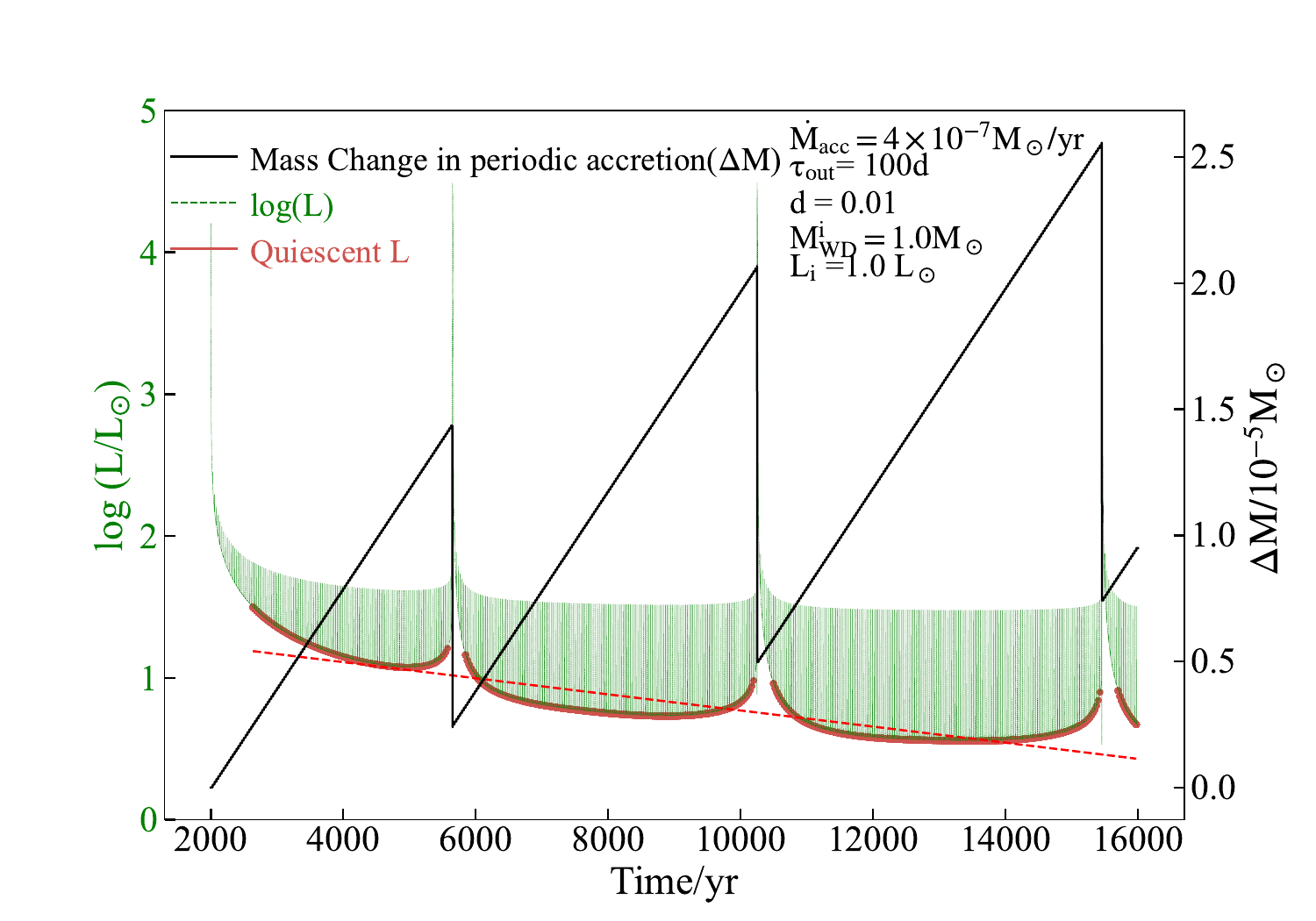} 
    \centering
    \caption{Evolution of a 1.0 \(\rm M_\odot\) CO WD undergoing periodic accretion. The black solid line traces the WD's mass variation (\(\Delta M\)), while the green dashed line shows evolution of luminosity (log\(L\)); Prior to the periodic accretion, the system sustained stable H-rich accretion for 2000 years (not shown in the figure). Then, the WD undergoes a periodic accretion phase during which the mass accretion rate is \(4 \times 10^{-7} \, \rm M_{\odot}/\text{yr}\) with a duty cycle of 0.01. The red solid line represents the quiescent state luminosity, while the red dashed line illustrates its declining trend over time.} 
    \label{fig:LM} 
\end{figure}

Our simulations show that periodic accretion in DNe cannot sustain stable H burning. Instead, the accumulated material eventually triggers nova eruptions. Fig.~\ref{fig:LM} illustrates the luminosity and mass evolution of a 1 \(\rm M_{\odot} \) CO WD undergoing periodic accretion at a rate of \( 4 \times 10^{-7}\, \rm M_{\odot}/\text{yr} \). The black solid line represents the WD mass variation (\(\Delta M\)) during the periodic accretion, whereas the green dashed line denotes the evolution of the WD's luminosity \(L\). After 2000 yr of stable accretion, the system enters a 10,000-day quiescent phase, during which mass accretion ceases entirely. The quiescent period is followed by a 100-day outburst phase, in which material is accreted onto the WD at a rate of \(\dot M_{\rm acc} = 4 \times 10^{-7}\, \rm M_{\odot}/\text{yr} \). This accretion cycle repeats periodically.

In Fig.~\ref{fig:LM}, the numerous DN outbursts manifest as dense, small-scale luminosity variations that appear as a blurred green band. During these outbursts, the luminosity enhancements are dominated by the release of gravitational potential energy associated with episodic accretion onto the WD (\citealp{townsley2004,piro2005,townsley2009}). As mass accumulates in the accreted envelope, the density at the base of the shell gradually increases. At approximately 5600 years, the temperature and density at the shell base reach the critical conditions required for thermonuclear runaway hydrogen burning, leading to a nova eruption. The eruption induces a super-Eddington wind (\(\eta_{\text{Edd}} = 1\)), expelling a significant fraction of the previously accumulated mass. After the nova explosion, the system resumes its periodic accretion pattern. At this stage, the WD interior is able to lose energy, resulting in a cooler and more degenerate CO core (\citealp{yaron2005}). Consequently, a longer period of mass accumulation is required before the second nova explosion. As a result, the second nova ejects a larger amount of material. This evolutionary trend becomes more pronounced in subsequent cycles. Furthermore, the cooling effect leads to a progressive decline in the quiescent luminosity, as illustrated by the red dashed line in Fig.~\ref{fig:LM}. 

\begin{figure}[htbp]  
    \includegraphics[width=0.45\textwidth]{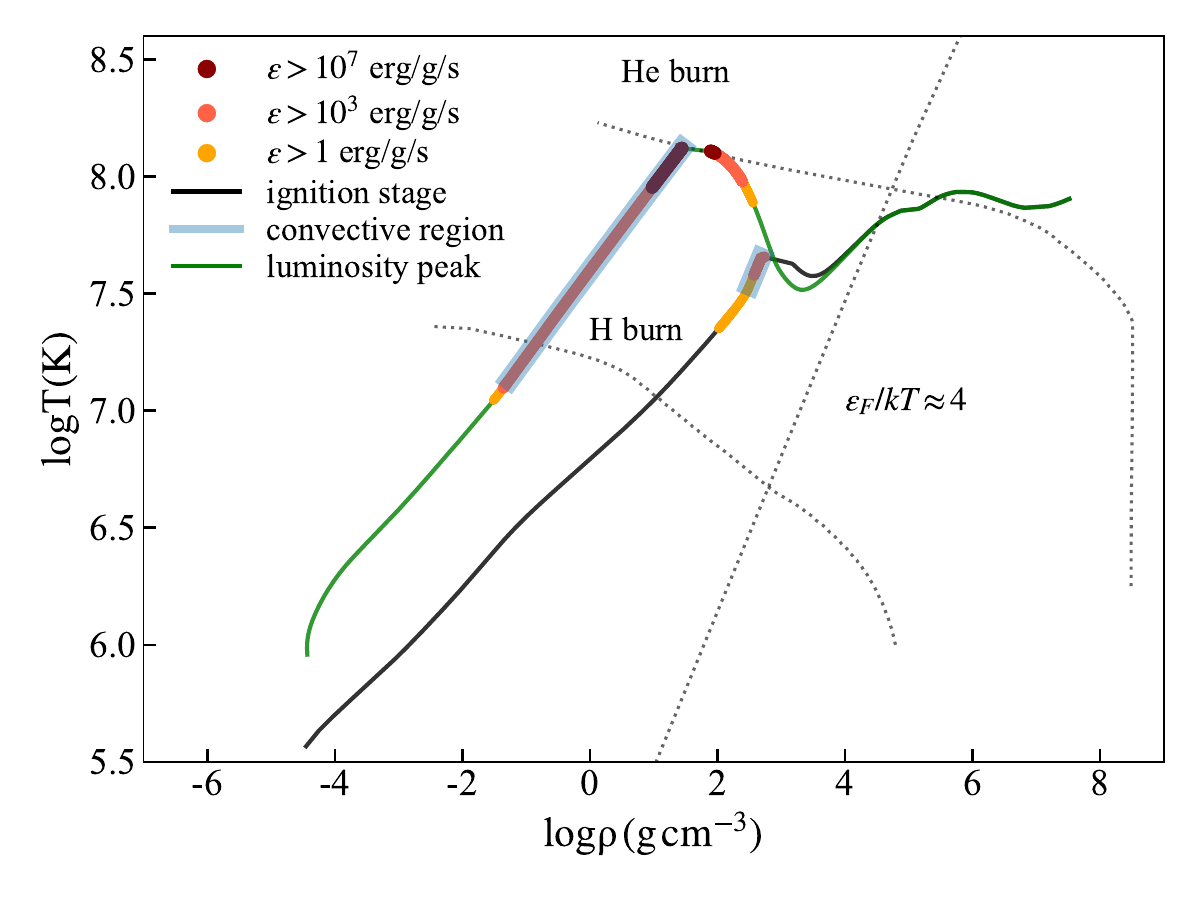} 
    \centering
    \caption{The same system as in Fig.~\ref{fig:LM}, but showing the density-temperature profiles during the first nova outburst. The black solid line represents the profile at the onset of H ignition, while the green solid line corresponds to the profile at the luminosity peak. Semitransparent blue indicates convective zones. In the burning zone, the burgundy-to-yellow gradients represent the energy generation rates (high to low). The H- and He-burning boundaries are based on tabulated reference curves provided with MESA and used in its standard plotting routines (\citealt{paxton2010modules}).}
    \label{fig:rhoT} 
\end{figure}

\begin{figure}[htbp]
    \includegraphics[width=0.45\textwidth]{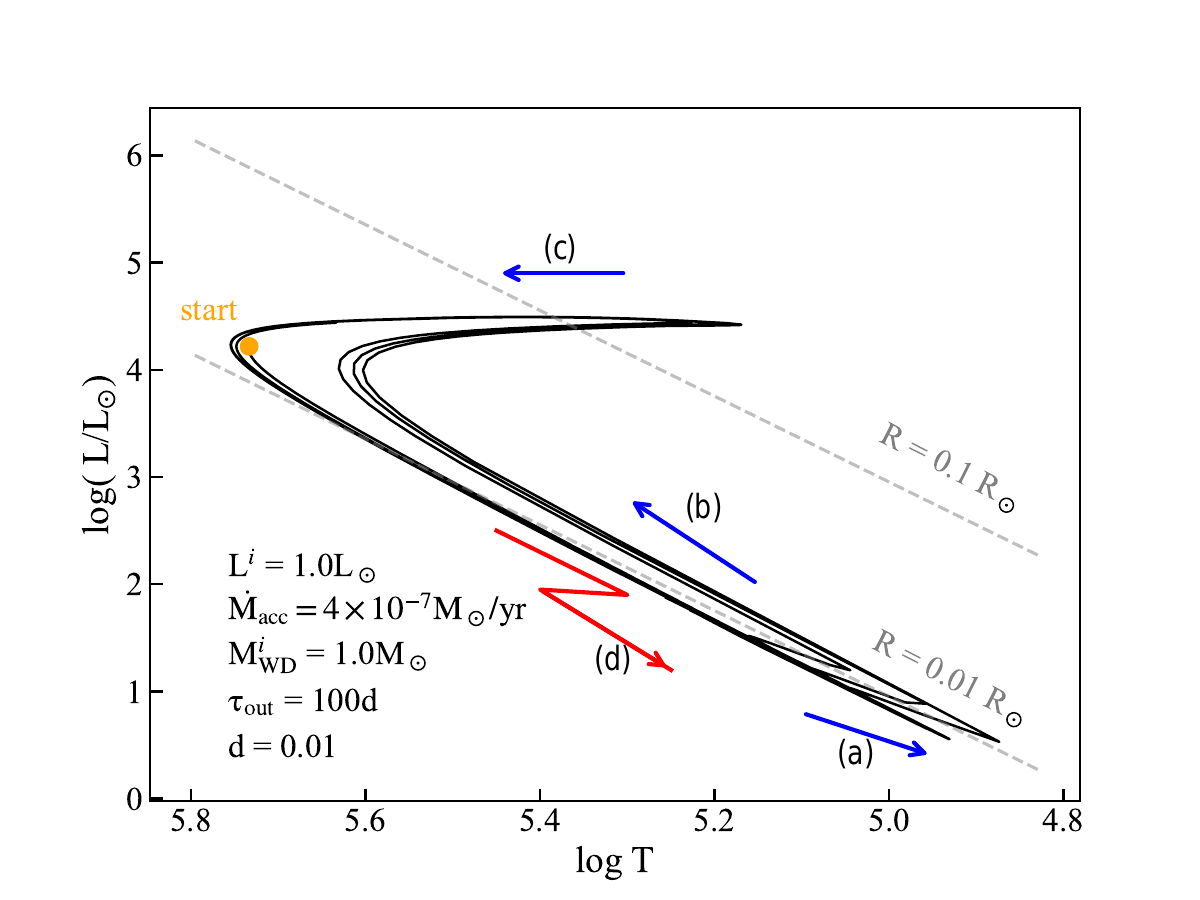} 
    \centering
    \caption{The same system as in Fig.~\ref{fig:LM}, but for the Hertzsprung-Russell diagram. The yellow filled circle marks the onset of periodic accretion, while the preceding 2000 years of stable accretion are not included. Blue arrows mark the nova outburst phases, and red arrow indicate periodic accretion phase, schematically illustrating the approximate luminosity–temperature variations during this stage.}
    \label{fig:HR} 
\end{figure}

Fig.~\ref{fig:rhoT} presents the temperature-density profile at both the onset of H ignition and the peak luminosity of the first nova outburst. In our models, the formation of a convective zone indicates the ignition of H burning. Ignition begins at the base of the shell, and within 200 days, thermal energy is transported to the surface through convective expansion. In Fig.~\ref{fig:HR}, we present the Hertzsprung-Russell diagram for the same system as in Fig.~\ref{fig:LM}. The yellow filled circle marks the onset of periodic accretion. Three nova eruption episodes occur during the periodic accretion phase, each characterized by three distinct stages (annotated by blue arrows in the diagram): (a) The ignited H shell expands, causing a decline in surface temperature (\(T_{\text{eff}}\)) and luminosity (\(L\)) as the radius increases. (b) Efficient energy transport via convection rapidly increases the radiative luminosity. This phase culminates in \(L\) exceeding the Eddington luminosity limit (\(L_{\text{Edd}}\)). (c) \(L\) exceeds \(L_{\text{Edd}}\), triggering the super-Eddington wind to lose matter. These winds gradually subside as the expanding photosphere reduces radiative driving efficiency, ultimately ceasing when the luminosity returns to sub-Eddington levels, thereby concluding the nova eruption cycle.

Following each nova, the system returns to the periodic accretion mode. During accretion, the addition of H-rich material onto the WD surface compresses the outer layers, releasing gravitational potential energy that heats the surface and increases the luminosity \(L\); conversely, during quiescence, the absence of accretion allows the outer layers to cool, causing \(L\) to decline. As a result, the luminosity oscillates within the region indicated by the red arrow in Fig.~\ref{fig:HR}(d). During this phase, H-rich material gradually accumulates on the WD surface until it reaches critical ignition conditions, setting the stage for the next nova eruption. Notably, the minimum \(T_{\text{eff}}\) and \(L\) of each nova explosion decrease over time due to the gradual cooling of the CO core.

We investigate how the initial WD temperature, accretion rate, accretion timescale, and quiescent timescale during the DN phase affect the system’s evolution. In our approach, one parameter is varied at a time while the others remain fixed, allowing us to isolate each effect.

\begin{figure*}[ht]
\centering

\begin{minipage}{0.46\textwidth}
\centering
\includegraphics[width=\textwidth]{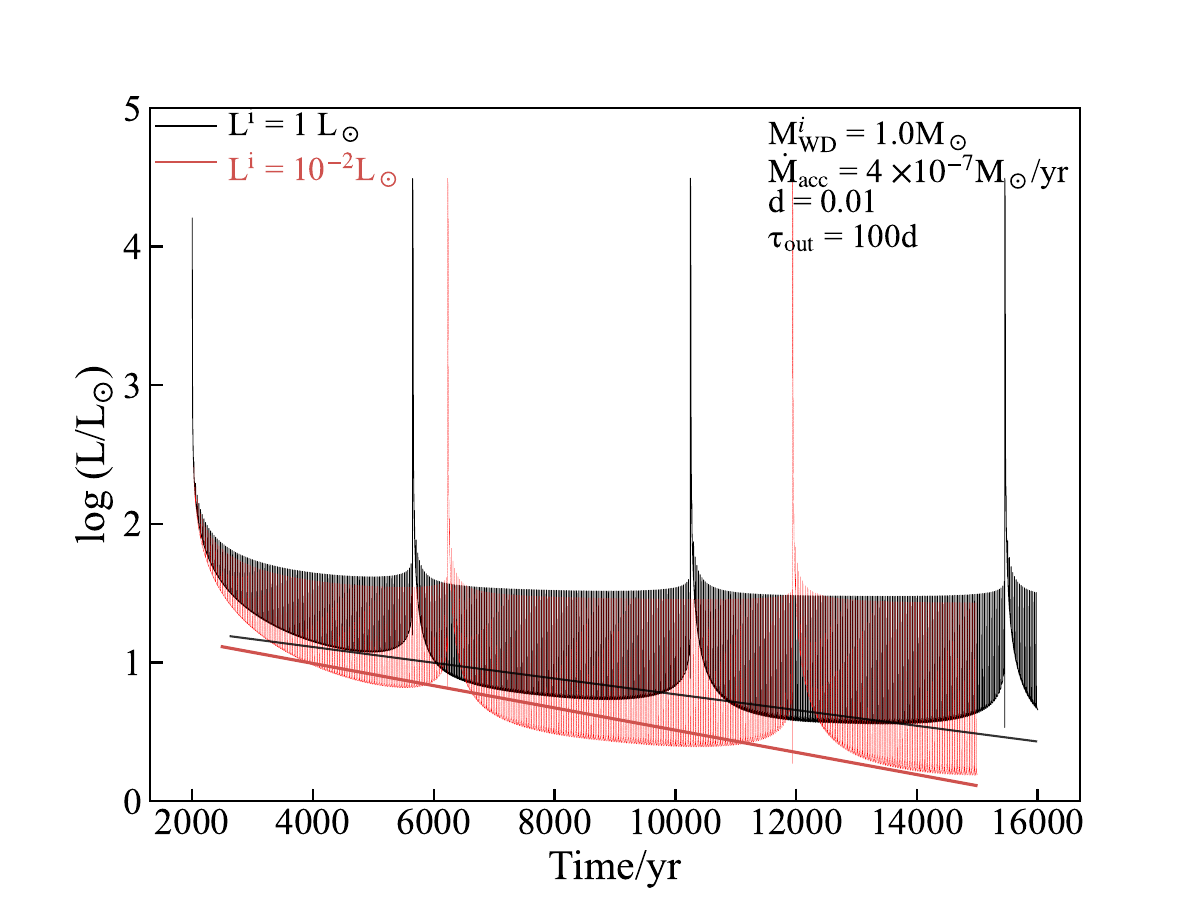}
(a)
\end{minipage}
\hfill
\begin{minipage}{0.46\textwidth}
\centering
\includegraphics[width=\textwidth]{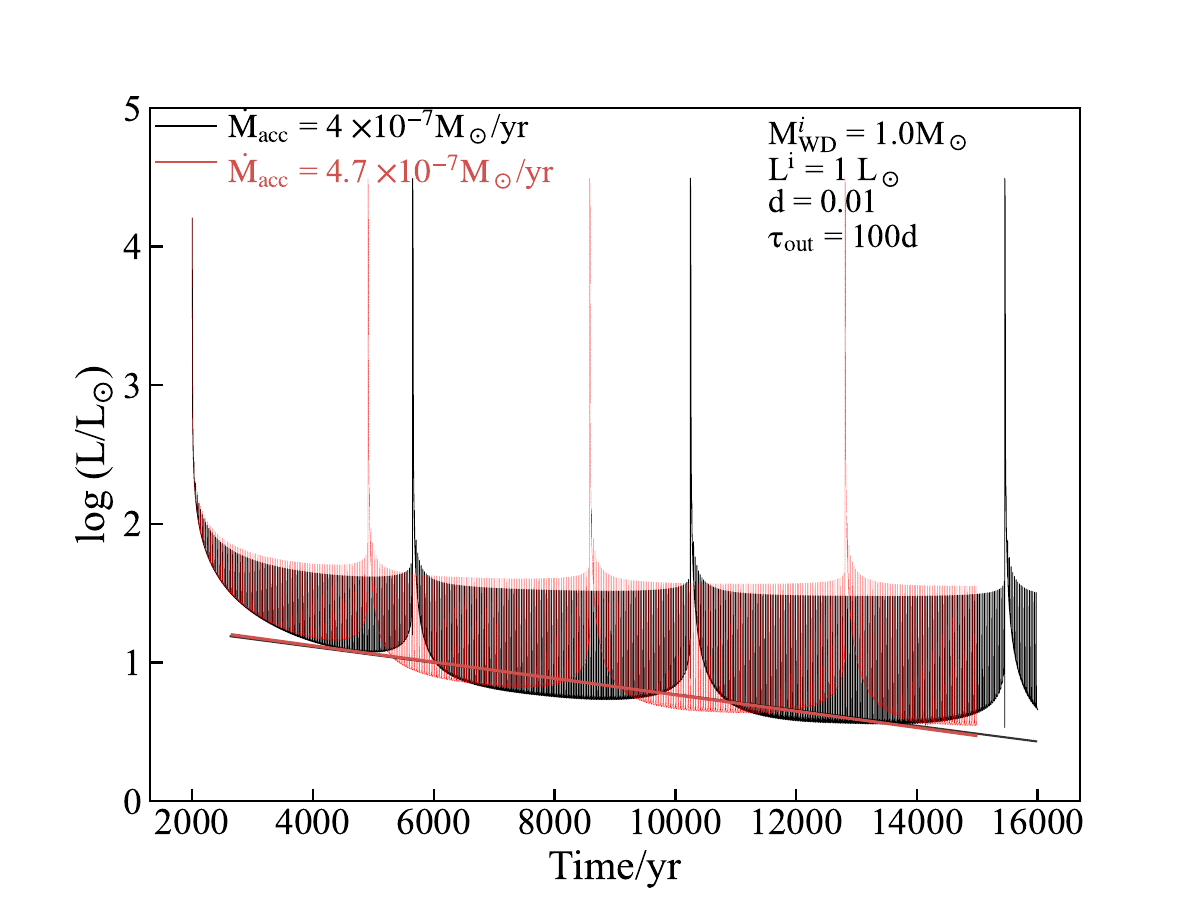}
(b)
\end{minipage}

\begin{minipage}{0.46\textwidth}
\centering
\includegraphics[width=\textwidth]{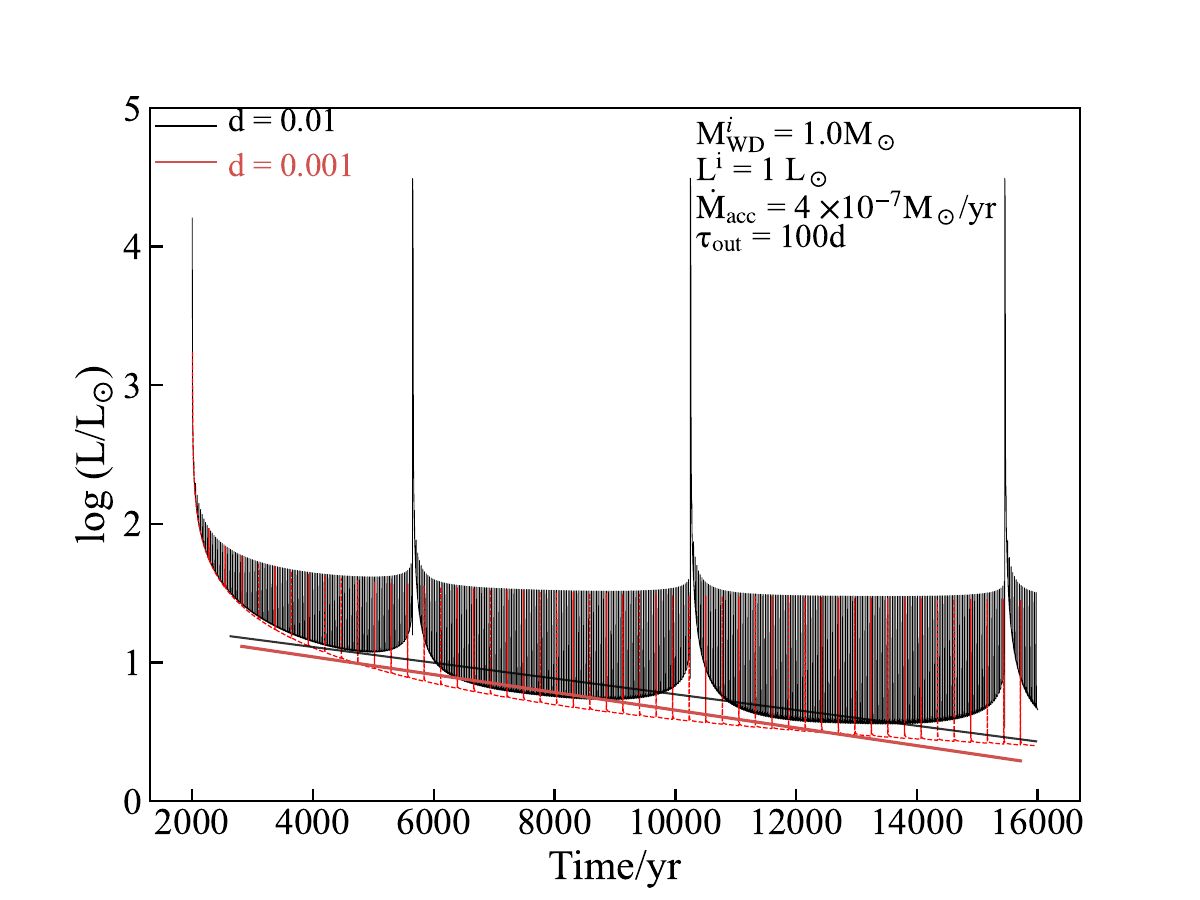}
(c)
\end{minipage}
\hfill
\begin{minipage}{0.46\textwidth}
\centering
\includegraphics[width=\textwidth]{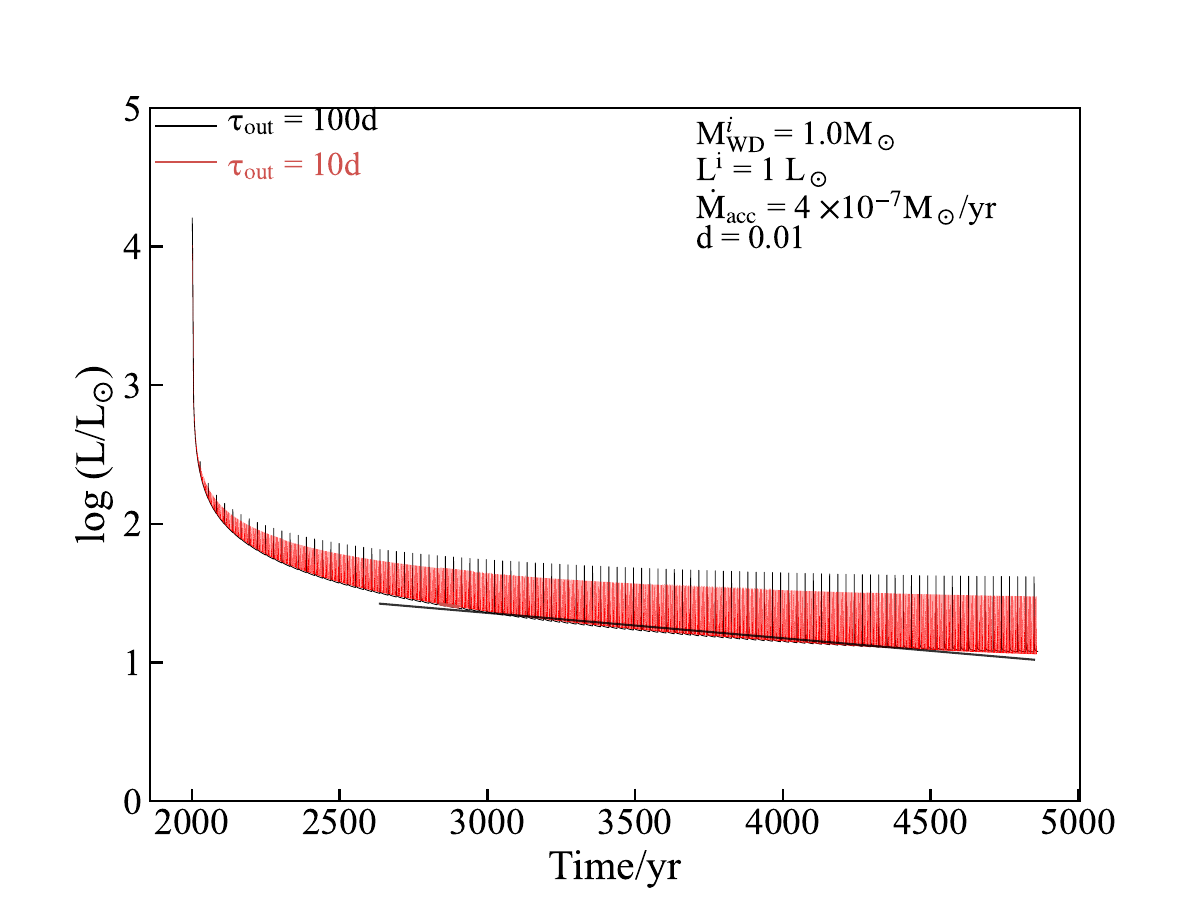}
(d)
\end{minipage}

\caption{
Combined effects of various parameters on the evolution of a CO white dwarf undergoing periodic accretion. Panel (a) shows the effect of the initial WD temperature, where the black line corresponds to an initial luminosity of $1\,\rm L_\odot$ and the red line to $10^{-2}\,\rm L_\odot$. Panel (b) shows the effect of the accretion rate, with the black line representing $\dot{M}_{\rm acc} = 4\times10^{-7}\,\rm M_\odot\,yr^{-1}$ and the red line $4.7\times10^{-7}\,\rm M_\odot\,yr^{-1}$. Panel (c) shows the effect of the duty cycle, where the black line represents $d=0.01$ and the red line $d=0.001$. Panel (d) shows the effect of the outburst duration, with the black line corresponding to $\tau_{\rm out}=100\,\mathrm{d}$ and the red line to $\tau_{\rm out}=10\,\mathrm{d}$.
}

\label{fig:composite}
\end{figure*}

To assess the influence of the initial WD temperature, we construct a CO WD with an initial luminosity of \(10^{-2}\, \rm L_{\odot}\), corresponding to an initial center temperature \(\log \left( {T_c}/{\mathrm{K}} \right) \approx 7.2\), compared to the standard model with \(\log \left( {T_c}/{\mathrm{K}} \right) \approx 7.9\). All other parameters are kept identical. As illustrated in panel (a) of Fig.~\ref{fig:composite}, a lower initial WD temperature results in a more rapid decline in quiescent-phase luminosity. This is because the different initial WD temperatures lead to distinct thermal structures at the onset of the periodic accretion phase. During the initial 2000 yr of steady accretion, the cooler WD requires a longer time to trigger the super-Eddington wind, allowing more H-rich material to accumulate and be subsequently ejected. As a result, by the end of this phase, the thermal profiles of the two models differ. At the onset of periodic accretion, although both models reach a similar peak temperature of $\log (T_{\mathrm{peak}}/\mathrm{K}) \approx 7.9$, this peak occurs at slightly different densities ($\log (\rho/\mathrm{g\,cm^{-3}}) \approx 4.2$ and 4.1 for the $L=1\,\rm L_{\odot}$ and $10^{-2}\,\rm L_{\odot}$ cases, respectively). The lower-density burning region in the cooler WD implies weaker degeneracy and a more efficient thermal response of the envelope, allowing the deposited heat to be radiated away more efficiently, and thus resulting in a more rapid decline of the luminosity during the quiescent phase.

To evaluate the impact of the accretion rate, we increase it during the periodic accretion phase to the upper limit of the stable burning regime. The results presented in panel (b) of Fig.~\ref{fig:composite} demonstrate that a higher accretion rate promotes a faster accumulation of H-rich material on the WD surface, thereby accelerating the onset of thermal runaway due to the increased mass accreted within the same period.

Observational studies suggest that DN typically exhibit long quiescent timescales. Assuming a constant outburst duration, this corresponds to a smaller duty cycle. To explore this effect, we set \( d = 0.001 \). As shown in panel (c) of Fig.~\ref{fig:composite}, while the overall decline in quiescent luminosity follows a similar trend, the mass accumulation timescale increases significantly, with the first nova eruption occurring after approximately \(10^5\) years.

Additionally, since observed outbursts are generally short-lived, we attempt to model an outburst duration of 10 days. However, due to numerical constraints, achieving such a short timescale requires significantly smaller time step, making long-term simulations computationally demanding. Consequently, we simulate this scenario over a 5000-year period, with the results displayed in panel (d) of Fig.~\ref{fig:composite}. These findings suggest that the long-term decline in quiescent luminosity is primarily governed by the WD’s core temperature, whereas variations in accretion parameters (such as accretion rate, outburst duration, and duty cycle) have a small effect on this trend.

Table \ref{tab:summary} summarizes the nova eruption parameters and outcomes for several models with an initial WD mass of \(1.0\, \rm M_{\odot}\). The table lists the initial luminosity (\(L^i\)), duty cycle (\(d\)), accretion rate (\(\dot{M}_{\text{acc}}\)), and outburst timescale (\(\tau_{\text{out}}\)). For models with multiple nova eruptions, the “Nova \#” column indicates the order of the eruptions (i.e., 1 = first eruption, 2 = second eruption, 3 = third eruption, etc.), and the outcome parameters include the evolutionary timescale reached at the time of the nova eruption, the mass accumulation efficiency (\(\eta_{\text{acc}}\)), and the envelope mass (\(M_{\text{env}}\)) for each event. The results show that increasing the accretion rate effectively enhances the mass accumulation efficiency; however, even when raised to the upper limit of the stable burning regime for H-rich material, it still fails to enable the stable H burning. The WD with lower \(L^i\) exhibit reduced \(\eta_{\text{acc}}\) in subsequent nova outbursts. A decrease in the d also leads to lower \(\eta_{\text{acc}}\). Moreover, the \(\eta_{\text{acc}}\) declines with each nova eruption because even though eruptions raise the WD temperature, this warmth is lost to radiative cooling during quiescence. Consequently, the CO core cools and becomes increasingly degenerate, necessitating the accumulation of a thicker H-rich envelope before each eruption, which in turn results in more violent outbursts.

\begin{table*}[ht]
\caption{Comparison of nova eruption parameters and outcomes for models with
$M^i_{\rm WD}=1.0\,\rm M_\odot$.}
\label{tab:summary}
\centering
\begin{tabular}{cccccccccc}
\hline
Set &
$L^i$ &
$d$ &
$\dot{M}_{\rm acc}$ &
$\tau_{\rm out}$ &
Nova \# &
$t$ &
$\eta_{\rm acc}$ &
$M_{\rm env}$ \\
 &
($L_\odot$) &
 &
($\rm M_\odot\,yr^{-1}$) &
(days) &
 &
(year) &
 &
($10^{-3}\,\rm M_\odot$) \\
\hline
1 & 1 & 0.01 & $4\times10^{-7}$ & 100 & 1 & 5648  & 0.167 & 2.071 \\
  &   &      &                  &     & 2 & 10247 & 0.141 & 2.079 \\
  &   &      &                  &     & 3 & 15454 & 0.118 & 2.083 \\

2 & $10^{-2}$ & 0.01 & $4\times10^{-7}$ & 100 & 1 & 6231  & 0.114 & 3.404 \\
  &           &      &                  &     & 2 & 11940 & 0.090 & 3.411 \\

3 & 1 & 0.01 & $4.7\times10^{-7}$ & 100 & 1 & 4910  & 0.181 & 2.071 \\
  &   &      &                    &     & 2 & 8582  & 0.146 & 2.077 \\
  &   &      &                    &     & 3 & 12809 & 0.131 & 2.082 \\

4 & 1 & 0.01 & $4\times10^{-7}$ & 10  & -- & -- & -- & -- \\

5 & 1 & 0.001 & $4\times10^{-7}$ & 100 & 1 & 66697 & 0.075 & 2.088 \\
\hline
\end{tabular}

\tablefoot{
Table~\ref{tab:summary} lists the nova eruption parameters and outcomes for
models with an initial WD mass of $1.0\,\rm M_\odot$.
Columns give the initial luminosity ($L^i$), duty cycle ($d$),
accretion rate ($\dot{M}_{\rm acc}$), and outburst duration
($\tau_{\rm out}$).
For models with nova eruptions, we list the eruption time $t$,
the mass accumulation efficiency $\eta_{\rm acc}$,
and the envelope mass $M_{\rm env}$.
}
\end{table*}

\begin{figure}[htbp]
\centering

\includegraphics[width=0.95\linewidth]{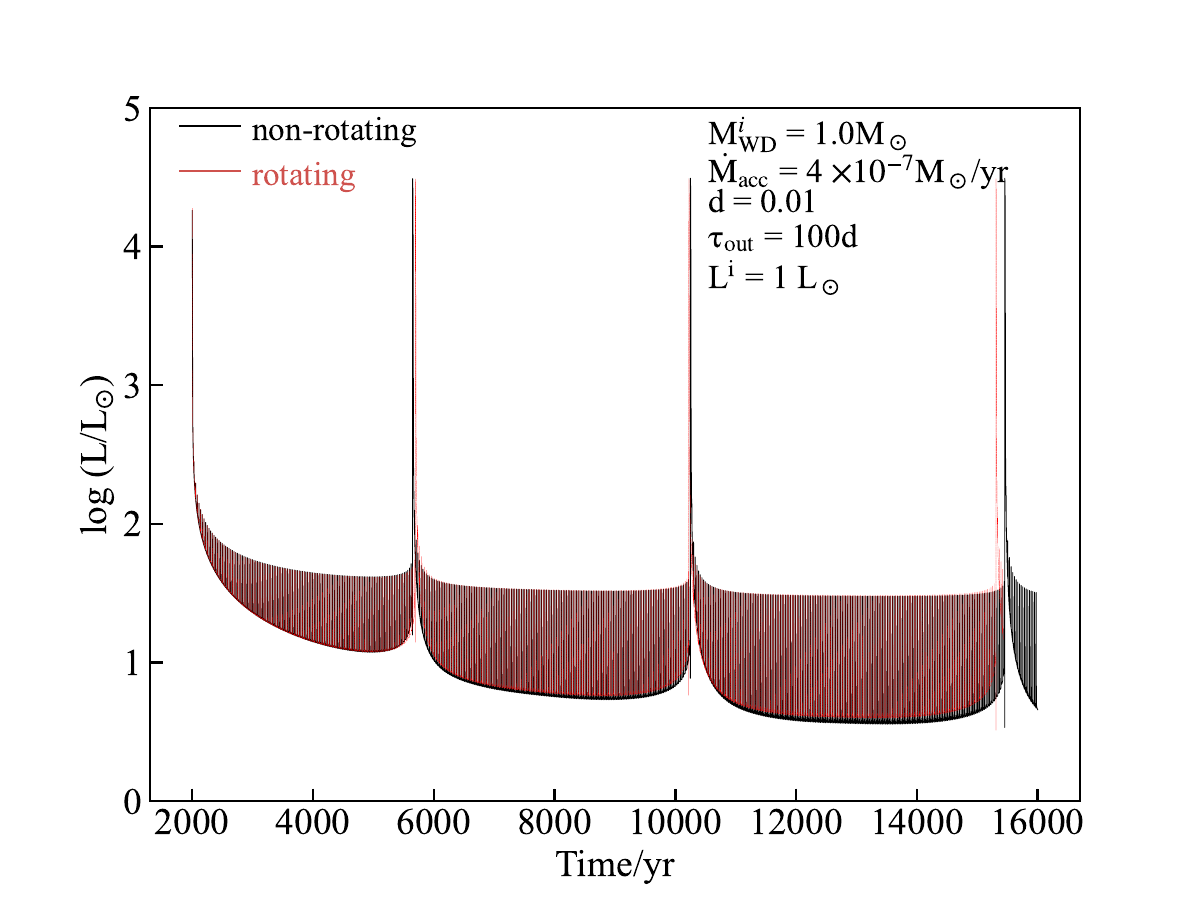}
(a)

\includegraphics[width=0.95\linewidth]{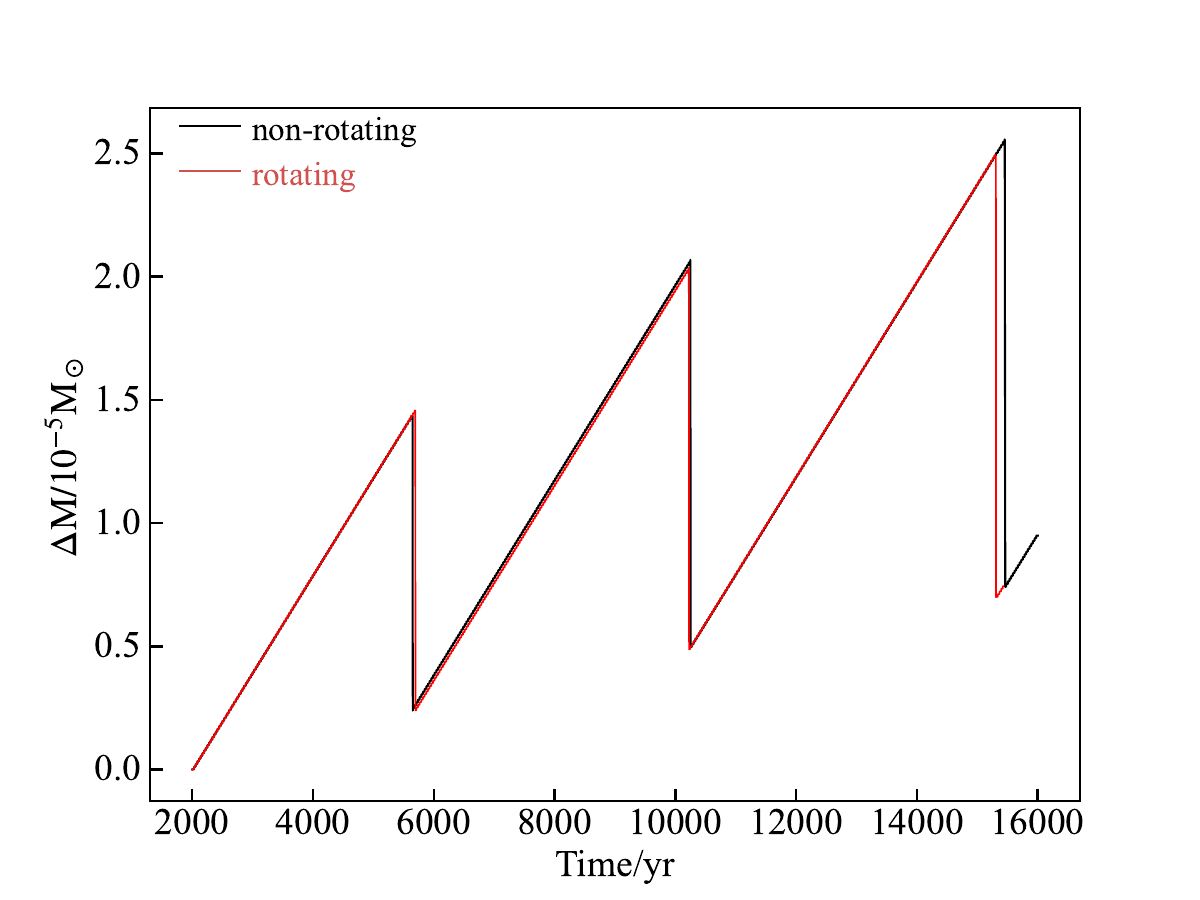}
(b)

\includegraphics[width=0.95\linewidth]{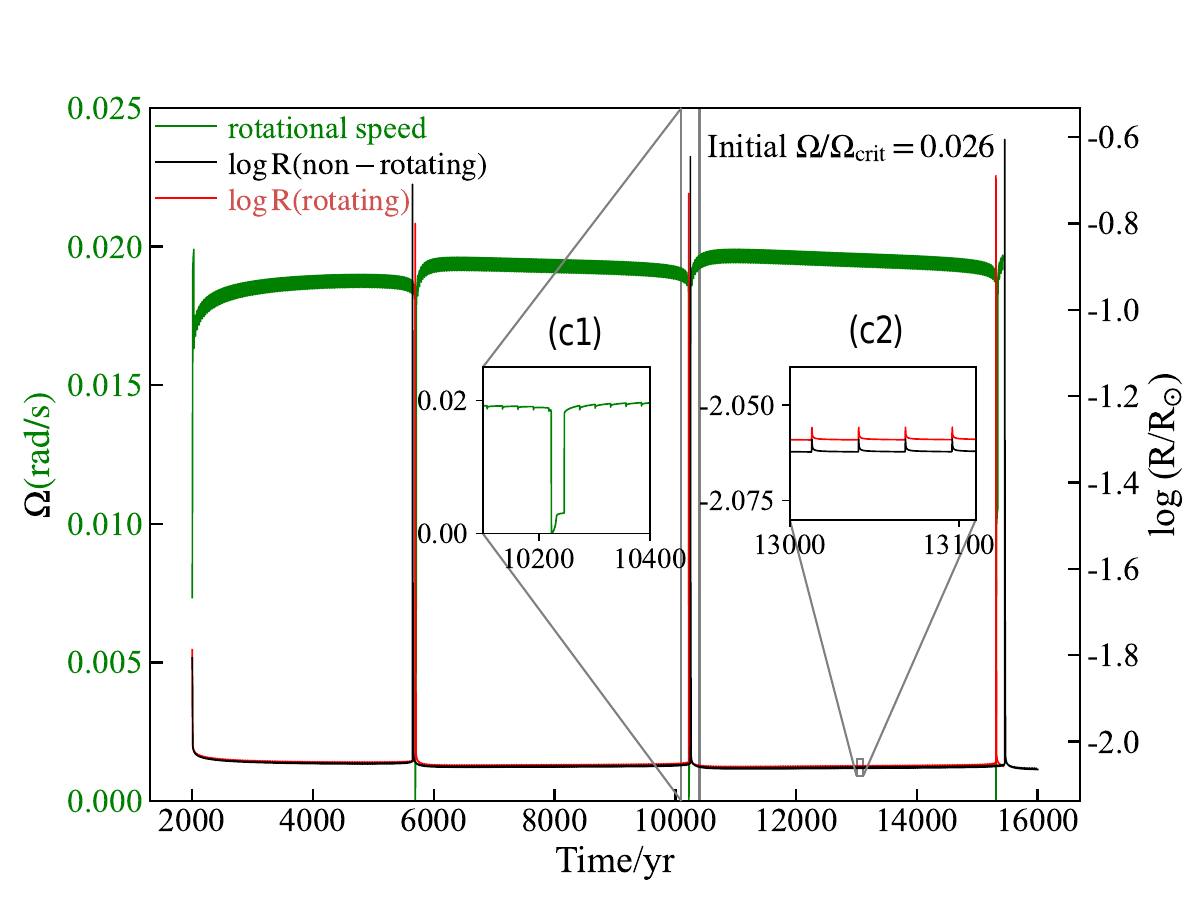}
(c)

\caption{
Comparison of the long-term evolution of a rotating and a non-rotating $1.0\,\rm M_\odot$ CO WD under periodic accretion. Panel (a) shows the evolution of the luminosity ($\log L$), panel (b) shows the evolution of the WD mass ($\Delta M$), and panel (c) shows the evolution of the rotational angular velocity ($\Omega$) and radius ($\log R$). The two inset panels provide zoomed-in views: (c1) highlights the rapid variations in $\Omega$ over the interval 10\,200--10\,400~yr, while (c2) shows the evolution of $\log R$ over the interval 13\,000--13\,100~yr. The initial 2000~yr of stable accretion are not shown.
}

\label{fig:spin}
\end{figure}

When accreted material from the disk settles onto the WD,  the accompanying spin angular momentum (AM) is input to change its rotation. To analyze the effect of rotation, we constructed a uniformly rotating WD model with an initial surface spin of  $\Omega \approx 0.002~{\rm rad~s^{-1}}$ ($\Omega/\Omega_{\rm crit} = 0.026$, where $\Omega_{\rm crit}$ is the critical angular velocity at which the centrifugal force balances gravity at the stellar equator), while keeping other parameters identical to Set 1 in Table \ref{tab:summary}. The WD accreted H-rich material at the same accretion rate and specific AM. After 2000 years of stable accretion, the WD reaches a surface spin rate of $\Omega \approx 0.007~{\rm rad~s^{-1}}$ ($\Omega/\Omega_{\rm crit} = 0.045$). In our simulation, AM transport is treated as a diffusive process within MESA. We include the effects of Eddington–Sweet circulation and dynamical shear instability, as the latter plays a key role in transporting AM in the highly degenerate core, while the former dominates in the non-degenerate envelope of WD (\citealt{yoon2004}). As shown in Fig.~\ref{fig:spin}(a), the rotating WD still undergoes nova outbursts during periodic accretion. The first outburst occurs slightly later than that in Set~1, which can be attributed to the reduction in effective gravity caused by rotation. This lowers the pressure at the base of the accreted envelope, requiring a somewhat larger ignition mass and thus delaying the onset of the first outburst. In contrast, the third outburst in the rotating model is triggered slightly earlier. As the nova cycles proceed, rotation gradually modifies the thermal structure of the accreted envelope, leading to a shallower pressure gradient and slower cooling, so that the ignition conditions for the third outburst are reached earlier. As shown in Fig.~\ref{fig:spin}(b), the mass evolution of the rotating WD is nearly identical to that of Set~1. The accumulation efficiency of the first nova outburst is 0.165, while the subsequent two novae reach efficiencies of 0.138 and 0.105, respectively. Although a decreasing trend is still present, the overall retention efficiency in the rotating model is slightly lower than that in the non-rotating case, because the reduction in effective gravity makes it easier for material to be ejected. The green line in Fig.~\ref{fig:spin}(c) shows the evolution of the WD’s spin velocity: during quiescent phases, the rotation accelerates as the radius contracts, while during outbursts it decreases as the envelope expands. The radius undergoes rapid expansion during nova eruptions, leading to a significant decrease in the spin rate. After the third nova outburst, the rotation rate has increased noticeably compared to its initial value, indicating a long-term spin-up trend. Overall, our results suggest that, even with rotation, the net mass accumulation remains inefficient over repeated nova cycles, and the WD is unlikely to reach the \(M_{\rm Ch}\) over long-term evolution.

\section{Discussions and conclusions} \label{sec:discussion}
According to \citet{king2003new}, DNe could theoretically accrete sufficient mass to increase \(1 \rm M_\odot\) WD to \(M_{\rm Ch}\). However, our simulations fail to produce SNe Ia because the periodic accretion pattern cannot sustain stable H burning. In Set 1, the equivalent mass transfer rate \(\dot{M}_2 = \dot{M}_{\text{acc}} \times d = 4.0 \times 10^{-9}\,\rm M_\odot/\text{yr}\). Assuming a linearly decreasing \(\eta_{\rm acc}\), this efficiency would approach zero after roughly \(6 \times 10^5\) years. During this period, approximately \(2.4 \times 10^{-3}\,\rm M_\odot\) of H-rich material is accreted, but only \(2 \times 10^{-4}\,\rm M_\odot\) may remain on the WD surface. This amount of material is insufficient to approach $M_{\rm Ch}$. Set 3 exhibits the highest mass accumulation efficiency. Nevertheless, its \(\eta_{\rm acc}\) also trends toward zero after \(5 \times 10^5\) years, with \(\dot{M}_2 = 4.7 \times 10^{-9}\,\rm M_\odot/\text{yr}\). Despite accreting \(2.4 \times 10^{-3}\,\rm M_\odot\) of H-rich material (with an average \(\eta_{\rm acc}\sim\) 0.09), only \(2 \times 10^{-4}\,\rm M_\odot\) likely persists on the WD surface, again failing to reach $M_{\rm Ch}$. While \(\eta_{\rm acc}\) is strongly dependent on the input stellar wind model, its secular decline remains a universal feature (since the pronounced cooling effect drives the temperature reduction of the CO core).

In DNe, WDs are generally more massive than isolated WDs, with some up to 1.2 \(\rm M_\odot\)\citep{godon2006hubble}. The mass ratio of the system is typically \(q \lesssim 1\). Based on our simulations, assuming an extreme case of a $1.2\,\rm M_\odot$ WD and a $1.2 \, \rm M_\odot$ MS donor system with duty cycle of $d = 0.1$, the maximum mass transfer rate that keeps the accretion rate within the H stable burning regime is $\dot{M}_2 = 6.2 \times 10^{-8}\,\rm M_\odot/\text{yr}$. At this mass transfer rate, complete mass transfer from the donor to the WD would occur over $\sim 2 \times 10^7$ years. However, achieving a SN Ia before the donor's mass is fully transferred requires an average accretion efficiency $\eta_{\text{acc}} > 0.148$ due to the mass loss caused by nova outbursts. However, in practice, the core of the WD is steadily cooling, leading to a gradual reduction in the accretion efficiency for nova outbursts; moreover, the exhaustion of the donor's H-rich envelope terminates mass transfer. Therefore, DNe are highly unlikely to produce SNe Ia.
 
In addition, \citet{king2003new} proposed that the duty cycle $d$ typically ranges between 0.1 and 0.001. In our simulations, we did not explore larger values of $d$, because a larger duty cycle corresponds to a shorter quiescent phase, which leads to more frequent accretion cycles within the same evolutionary timescale. This significantly increases the computational cost. Therefore, we only considered cases with $d = 0.001$ and $d = 0.01$. Theoretically, smaller values of \(d\) should enhance the equivalent accretion rate, thus promoting the growth of the WD mass. However, as seen in Sets 1 and 5, reduced $d$ extends the cooling time of the WD, leading to easier nova outbursts and a lower likelihood of producing SNe Ia. For larger \(d\), considering that the outburst timescale of DN is on the order of day, the short quiescent state limits the accumulated mass of each cycle. Even if sufficient mass is retained, the equivalent accretion rate during the outburst may not reach the stable H-burning regime. Our results suggest that even for very small \(d\), H shell flashes still occur, leading to recurrent nova outbursts and inefficient mass retention. This differs from \cite{liu2023type}, who proposed that very small \(d\) may trigger optically thick winds, while larger \(d\) leads to H shell flashes that eject most of the accreted material. Moreover, considering only \(d\) is insufficient because $d$ represents the ratio of the outburst to quiescent timescales. Even if $d$ remains constant, variations in either the outburst or quiescent durations may lead to different evolutionary outcomes.  

Some observations have found nova shells around DN \citep{shara2007ancient,shara2012cnc}, and many CVs have been shown to transition between novae and DNe \citep{GKPer2017, kato2021v606, murphy2022v392}, meaning that at least some novae will change to DNe. Our simulations are consistent with the established picture that repeated DN accretion outbursts can gradually accumulate sufficient mass to trigger nova eruptions.

In this study, we simulated the periodic accretion mode in DNe systems. However, uncertainty remains as to whether the entire disk matter can immediately settle onto the WD. Moreover, our model does not incorporate binary evolution; changes in the binary orbital period and the accretion rate could alter the accretion mode. Additionally, our simulations employed a simplified super-Eddington wind prescription to calculate mass ejection. However, The mass retention efficiency during nova eruptions is highly sensitive to stellar wind assumption; for example, slower winds may allow a portion of the ejected material to fall back onto the WD surface. These factors may introduce additional uncertainties into our results.

Using MESA, we simulated the evolution of a 1.0 $\rm M_\odot$ CO WD via periodic accretion. Our main results are summarized as follows: (i) H-rich material cannot burn stably during periodic accretion; degeneracy in long quiescence triggers nova eruptions. (ii) The temperature that increases during DNe outburst phase cools more during the quiescence phase, resulting in a gradual decay in luminosity during the quiescence phase. (iii) The decline in mass accumulation efficiency in nova cycles prevents WD from reaching $M_{\rm Ch}$, making DNe unlikely to be progenitors of SNe Ia. In the future, more observational identification and theoretical simulations of DN systems are needed.

\begin{acknowledgements}
We thank Prof. Xiangdong Li for his helpful discussions. This study is supported by the National Natural Science Foundation of China (Nos 12225304, 12288102 and 12473032), the CAS Project for Young Scientists in Basic Research (YSBR-148), the Yunnan Revitalization Talent Support Program (Young Talent project and Yunling Scholar Project), the Yunnan Science and Technology Program (Nos 202501AW070001, 202501AS070005, 202605AS350010 and 202601BC070011), and the International Centre of Supernovae (ICESUN), Yunnan Key Laboratory of Supernova Research (No. 202505AV340004).
     
\end{acknowledgements}
\bibliography{refs}

@ARTICLE{hoyle1960nucleosynthesis,
       author = {{Hoyle}, F. and {Fowler}, William A.},
        title = "{Nucleosynthesis in Supernovae.}",
      journal = {\apj},
         year = 1960,
        month = nov,
       volume = {132},
        pages = {565},
          doi = {10.1086/146963},
       adsurl = {https://ui.adsabs.harvard.edu/abs/1960ApJ...132..565H}
}

@ARTICLE{riess1998,
       author = {{Riess}, Adam G. and {Filippenko}, Alexei V. and {Challis}, Peter and {Clocchiatti}, Alejandro and {Diercks}, Alan and {Garnavich}, Peter M. and {Gilliland}, Ron L. and {Hogan}, Craig J. and {Jha}, Saurabh and {Kirshner}, Robert P. and {Leibundgut}, B. and {Phillips}, M.~M. and {Reiss}, David and {Schmidt}, Brian P. and {Schommer}, Robert A. and {Smith}, R. Chris and {Spyromilio}, J. and {Stubbs}, Christopher and {Suntzeff}, Nicholas B. and {Tonry}, John},
        title = "{Observational Evidence from Supernovae for an Accelerating Universe and a Cosmological Constant}",
      journal = {\aj},
         year = 1998,
        month = sep,
       volume = {116},
       number = {3},
        pages = {1009-1038},
          doi = {10.1086/300499},
archivePrefix = {arXiv},
       eprint = {astro-ph/9805201},
 primaryClass = {astro-ph},
       adsurl = {https://ui.adsabs.harvard.edu/abs/1998AJ....116.1009R}
}

@ARTICLE{perlmutter1999,
       author = {{Perlmutter}, S. and {Aldering}, G. and {Goldhaber}, G. and {Knop}, R.~A. and {Nugent}, P. and {Castro}, P.~G. and {Deustua}, S. and {Fabbro}, S. and {Goobar}, A. and {Groom}, D.~E. and {Hook}, I.~M. and {Kim}, A.~G. and {Kim}, M.~Y. and {Lee}, J.~C. and {Nunes}, N.~J. and {Pain}, R. and {Pennypacker}, C.~R. and {Quimby}, R. and {Lidman}, C. and {Ellis}, R.~S. and {Irwin}, M. and {McMahon}, R.~G. and {Ruiz-Lapuente}, P. and {Walton}, N. and {Schaefer}, B. and {Boyle}, B.~J. and {Filippenko}, A.~V. and {Matheson}, T. and {Fruchter}, A.~S. and {Panagia}, N. and {Newberg}, H.~J.~M. and {Couch}, W.~J. and {Project}, The Supernova Cosmology},
        title = "{Measurements of {\ensuremath{\Omega}} and {\ensuremath{\Lambda}} from 42 High-Redshift Supernovae}",
      journal = {\apj},
         year = 1999,
        month = jun,
       volume = {517},
       number = {2},
        pages = {565-586},
          doi = {10.1086/307221},
archivePrefix = {arXiv},
       eprint = {astro-ph/9812133},
 primaryClass = {astro-ph},
       adsurl = {https://ui.adsabs.harvard.edu/abs/1999ApJ...517..565P}
}

@ARTICLE{greggio1983,
       author = {{Greggio}, L. and {Renzini}, A.},
        title = "{The binary model for type I supernovae - Theoretical rates}",
      journal = {\aap},
         year = 1983,
        month = feb,
       volume = {118},
       number = {2},
        pages = {217-222},
       adsurl = {https://ui.adsabs.harvard.edu/abs/1983A&A...118..217G}
}

@ARTICLE{matteucci1986,
       author = {{Matteucci}, F. and {Greggio}, L.},
        title = "{Relative roles of type I and II supernovae in the chemical enrichment of the interstellar gas}",
      journal = {\aap},
         year = 1986,
        month = jan,
       volume = {154},
       number = {1-2},
        pages = {279-287},
       adsurl = {https://ui.adsabs.harvard.edu/abs/1986A&A...154..279M}
}

@ARTICLE{dwek2016iron,
       author = {{Dwek}, Eli},
        title = "{Iron: A Key Element for Understanding the Origin and Evolution of Interstellar Dust}",
      journal = {\apj},
         year = 2016,
        month = jul,
       volume = {825},
       number = {2},
          eid = {136},
        pages = {136},
          doi = {10.3847/0004-637X/825/2/136},
archivePrefix = {arXiv},
       eprint = {1605.01957},
 primaryClass = {astro-ph.GA},
       adsurl = {https://ui.adsabs.harvard.edu/abs/2016ApJ...825..136D}
}

@ARTICLE{hillebrandt2000type,
       author = {{Hillebrandt}, Wolfgang and {Niemeyer}, Jens C.},
        title = "{Type IA Supernova Explosion Models}",
      journal = {\araa},
         year = 2000,
        month = jan,
       volume = {38},
        pages = {191-230},
          doi = {10.1146/annurev.astro.38.1.191},
archivePrefix = {arXiv},
       eprint = {astro-ph/0006305},
 primaryClass = {astro-ph},
       adsurl = {https://ui.adsabs.harvard.edu/abs/2000ARA&A..38..191H}
}

@ARTICLE{roepke2005full,
       author = {{R{\"o}pke}, F.~K. and {Hillebrandt}, W.},
        title = "{Full-star type Ia supernova explosion models}",
      journal = {\aap},
         year = 2005,
        month = feb,
       volume = {431},
        pages = {635-645},
          doi = {10.1051/0004-6361:20041859},
archivePrefix = {arXiv},
       eprint = {astro-ph/0409286},
 primaryClass = {astro-ph},
       adsurl = {https://ui.adsabs.harvard.edu/abs/2005A&A...431..635R}
}

@ARTICLE{podsiadlowski2008nuclear,
       author = {{Podsiadlowski}, Philipp and {Mazzali}, Paolo and {Lesaffre}, Pierre and {Han}, Zhanwen and {F{\"o}rster}, Francisco},
        title = "{The nuclear diversity of Type Ia supernova explosions}",
      journal = {\nar},
         year = 2008,
        month = oct,
       volume = {52},
       number = {7-10},
        pages = {381-385},
          doi = {10.1016/j.newar.2008.06.020},
       adsurl = {https://ui.adsabs.harvard.edu/abs/2008NewAR..52..381P}
}

@ARTICLE{wang2012progenitors,
       author = {{Wang}, Bo and {Han}, Zhanwen},
        title = "{Progenitors of type Ia supernovae}",
      journal = {\nar},
         year = 2012,
        month = jun,
       volume = {56},
       number = {4},
        pages = {122-141},
          doi = {10.1016/j.newar.2012.04.001},
archivePrefix = {arXiv},
       eprint = {1204.1155},
 primaryClass = {astro-ph.SR},
       adsurl = {https://ui.adsabs.harvard.edu/abs/2012NewAR..56..122W}
}

@ARTICLE{liu2023type,
       author = {{Liu}, Zheng-Wei and {R{\"o}pke}, Friedrich K. and {Han}, Zhanwen},
        title = "{Type Ia Supernova Explosions in Binary Systems: A Review}",
      journal = {Research in Astronomy and Astrophysics},
         year = 2023,
        month = aug,
       volume = {23},
       number = {8},
          eid = {082001},
        pages = {082001},
          doi = {10.1088/1674-4527/acd89e},
archivePrefix = {arXiv},
       eprint = {2305.13305},
 primaryClass = {astro-ph.HE},
       adsurl = {https://ui.adsabs.harvard.edu/abs/2023RAA....23h2001L}
}

@ARTICLE{Ruiter2025A&ARv..33....1R,
       author = {{Ruiter}, Ashley Jade and {Seitenzahl}, Ivo Rolf},
        title = "{Type Ia supernova progenitors: a contemporary view of a long-standing puzzle}",
      journal = {\aapr},
         year = 2025,
        month = dec,
       volume = {33},
       number = {1},
          eid = {1},
        pages = {1},
          doi = {10.1007/s00159-024-00158-9},
archivePrefix = {arXiv},
       eprint = {2412.01766},
 primaryClass = {astro-ph.SR},
       adsurl = {https://ui.adsabs.harvard.edu/abs/2025A&ARv..33....1R}
}

@ARTICLE{Langer2000A&A...362.1046L,
       author = {{Langer}, N. and {Deutschmann}, A. and {Wellstein}, S. and {H{\"o}flich}, P.},
        title = "{The evolution of main sequence star + white dwarf binary systems towards Type Ia supernovae}",
      journal = {\aap},
         year = 2000,
        month = oct,
       volume = {362},
        pages = {1046-1064},
          doi = {10.48550/arXiv.astro-ph/0008444},
archivePrefix = {arXiv},
       eprint = {astro-ph/0008444},
 primaryClass = {astro-ph},
       adsurl = {https://ui.adsabs.harvard.edu/abs/2000A&A...362.1046L}
}

@ARTICLE{iben1984supernovae,
       author = {{Iben}, Jr., I. and {Tutukov}, A.~V.},
        title = "{Supernovae of type I as end products of the evolution of binaries with components of moderate initial mass.}",
      journal = {\apjs},
         year = 1984,
        month = feb,
       volume = {54},
        pages = {335-372},
          doi = {10.1086/190932},
       adsurl = {https://ui.adsabs.harvard.edu/abs/1984ApJS...54..335I}
}

@ARTICLE{webbink1984double,
       author = {{Webbink}, R.~F.},
        title = "{Double white dwarfs as progenitors of R Coronae Borealis stars and type I supernovae.}",
      journal = {\apj},
         year = 1984,
        month = feb,
       volume = {277},
        pages = {355-360},
          doi = {10.1086/161701},
       adsurl = {https://ui.adsabs.harvard.edu/abs/1984ApJ...277..355W}
}

@ARTICLE{wang2018mass,
       author = {{Wang}, Bo},
        title = "{Mass-accreting white dwarfs and type Ia supernovae}",
      journal = {Research in Astronomy and Astrophysics},
         year = 2018,
        month = may,
       volume = {18},
       number = {5},
          eid = {049},
        pages = {049},
          doi = {10.1088/1674-4527/18/5/49},
archivePrefix = {arXiv},
       eprint = {1801.04031},
 primaryClass = {astro-ph.SR},
       adsurl = {https://ui.adsabs.harvard.edu/abs/2018RAA....18...49W}
}

@ARTICLE{king2003new,
       author = {{King}, A.~R. and {Rolfe}, D.~J. and {Schenker}, K.},
        title = "{A new evolutionary channel for Type Ia supernovae}",
      journal = {\mnras},
         year = 2003,
        month = jun,
       volume = {341},
       number = {4},
        pages = {L35-L38},
          doi = {10.1046/j.1365-8711.2003.06639.x},
archivePrefix = {arXiv},
       eprint = {astro-ph/0303488},
 primaryClass = {astro-ph},
       adsurl = {https://ui.adsabs.harvard.edu/abs/2003MNRAS.341L..35K}
}

@ARTICLE{xu2009evolution,
       author = {{Xu}, Xiao-Jie and {Li}, Xiang-Dong},
        title = "{Evolution of long-period, white-dwarf binaries: application to GRO J1744-28 and type Ia supernovae}",
      journal = {\aap},
         year = 2009,
        month = feb,
       volume = {495},
       number = {1},
        pages = {243-248},
          doi = {10.1051/0004-6361:200810238},
       adsurl = {https://ui.adsabs.harvard.edu/abs/2009A&A...495..243X}
}

@ARTICLE{wang2010progenitors,
       author = {{Wang}, Bo and {Li}, Xiang-Dong and {Han}, Zhan-Wen},
        title = "{The progenitors of Type Ia supernovae with long delay times}",
      journal = {\mnras},
         year = 2010,
        month = feb,
       volume = {401},
       number = {4},
        pages = {2729-2738},
          doi = {10.1111/j.1365-2966.2009.15857.x},
archivePrefix = {arXiv},
       eprint = {0910.2138},
 primaryClass = {astro-ph.SR},
       adsurl = {https://ui.adsabs.harvard.edu/abs/2010MNRAS.401.2729W}
}

@ARTICLE{meng2010comprehensive,
       author = {{Meng}, X. and {Yang}, W.},
        title = "{A Comprehensive Progenitor Model for SNe Ia}",
      journal = {\apj},
         year = 2010,
        month = feb,
       volume = {710},
       number = {2},
        pages = {1310-1323},
          doi = {10.1088/0004-637X/710/2/1310},
archivePrefix = {arXiv},
       eprint = {0910.4992},
 primaryClass = {astro-ph.SR},
       adsurl = {https://ui.adsabs.harvard.edu/abs/2010ApJ...710.1310M}
}

@ARTICLE{saio1985evolution,
       author = {{Saio}, H. and {Nomoto}, K.},
        title = "{Evolution of a merging pair of C + O white dwarfs to form a single neutron star}",
      journal = {\aap},
         year = 1985,
        month = sep,
       volume = {150},
       number = {1},
        pages = {L21-L23},
       adsurl = {https://ui.adsabs.harvard.edu/abs/1985A&A...150L..21S}
}

@ARTICLE{saio1998inward,
       author = {{Saio}, Hideyuki and {Nomoto}, Ken'ichi},
        title = "{Inward Propagation of Nuclear-burning Shells in Merging C-O and He White Dwarfs}",
      journal = {\apj},
         year = 1998,
        month = jun,
       volume = {500},
       number = {1},
        pages = {388-397},
          doi = {10.1086/305696},
archivePrefix = {arXiv},
       eprint = {astro-ph/9801084},
 primaryClass = {astro-ph},
       adsurl = {https://ui.adsabs.harvard.edu/abs/1998ApJ...500..388S}
}

@ARTICLE{timmes1994conductive,
       author = {{Timmes}, F.~X. and {Woosley}, S.~E. and {Taam}, Ronald E.},
        title = "{The Conductive Propagation of Nuclear Flames. II. Convectively Bounded Flames in C+O and O+Ne+Mg Cores}",
      journal = {\apj},
         year = 1994,
        month = jan,
       volume = {420},
        pages = {348},
          doi = {10.1086/173565},
       adsurl = {https://ui.adsabs.harvard.edu/abs/1994ApJ...420..348T}
}

@INPROCEEDINGS{mochkovitch1997merging,
       author = {{Mochkovitch}, R. and {Guerrero}, J. and {Segretain}, L.},
        title = "{The merging of white dwarfs}",
    booktitle = {Thermonuclear Supernovae},
         year = 1997,
       editor = {{Ruiz-Lapuente}, P. and {Canal}, R. and {Isern}, J.},
       series = {NATO Advanced Study Institute (ASI) Series C},
       volume = {486},
        month = jan,
        pages = {187},
          doi = {10.1007/978-94-011-5710-0_13},
       adsurl = {https://ui.adsabs.harvard.edu/abs/1997ASIC..486..187M}
}

@ARTICLE{nomoto1984accreting,
       author = {{Nomoto}, K. and {Thielemann}, F.-K. and {Yokoi}, K.},
        title = "{Accreting white dwarf models for type I supernovae. III. Carbon deflagration supernovae.}",
      journal = {\apj},
         year = 1984,
        month = nov,
       volume = {286},
        pages = {644-658},
          doi = {10.1086/162639},
       adsurl = {https://ui.adsabs.harvard.edu/abs/1984ApJ...286..644N}
}

@ARTICLE{branch1985accreting,
       author = {{Branch}, D. and {Doggett}, J.~B. and {Nomoto}, K. and {Thielemann}, F.-K.},
        title = "{Accreting white dwarf models for type I supernovae. IV. The optical spectrum of a carbon-deflagration supernova.}",
      journal = {\apj},
         year = 1985,
        month = jul,
       volume = {294},
        pages = {619-625},
          doi = {10.1086/163329},
       adsurl = {https://ui.adsabs.harvard.edu/abs/1985ApJ...294..619B}
}

@ARTICLE{meyer1981elusive,
       author = {{Meyer}, F. and {Meyer-Hofmeister}, E.},
        title = "{On the elusive cause of cataclysmic variable outbursts.}",
      journal = {\aap},
         year = 1981,
        month = jan,
       volume = {104},
        pages = {L10-L12},
       adsurl = {https://ui.adsabs.harvard.edu/abs/1981A&A...104L..10M}
}

@INCOLLECTION{king1997disc,
       author = {{King}, A.~R.},
        title = "{Disc instabilities and binary evolution}",
    booktitle = {Accretion Disks - New Aspects},
         year = 1997,
       editor = {{Meyer-Hofmeister}, Emmi and {Spruit}, Henk},
       volume = {487},
        pages = {89},
          doi = {10.1007/BFb0105823},
       adsurl = {https://ui.adsabs.harvard.edu/abs/1997LNP...487...89K}
}

@ARTICLE{osaki1974accretion,
       author = {{Osaki}, Y.},
        title = "{An Accretion Model for the Outbursts of U Geminorum Stars}",
      journal = {\pasj},
         year = 1974,
        month = nov,
       volume = {26},
       number = {4},
        pages = {429-436},
          doi = {10.1093/pasj/26.4.429},
       adsurl = {https://ui.adsabs.harvard.edu/abs/1974PASJ...26..429O}
}

@ARTICLE{nomoto2007thermal,
       author = {{Nomoto}, Ken'ichi and {Saio}, Hideyuki and {Kato}, Mariko and {Hachisu}, Izumi},
        title = "{Thermal Stability of White Dwarfs Accreting Hydrogen-rich Matter and Progenitors of Type Ia Supernovae}",
      journal = {\apj},
         year = 2007,
        month = jul,
       volume = {663},
       number = {2},
        pages = {1269-1276},
          doi = {10.1086/518465},
archivePrefix = {arXiv},
       eprint = {astro-ph/0603351},
 primaryClass = {astro-ph},
       adsurl = {https://ui.adsabs.harvard.edu/abs/2007ApJ...663.1269N}
}

@software{paxton2010modules,
       author = {{Paxton}, Bill and {Bildsten}, Lars and {Dotter}, Aaron and {Herwig}, Falk and {Lesaffre}, Pierre and {Timmes}, Frank},
        title = "{MESA: Modules for Experiments in Stellar Astrophysics}",
 howpublished = {Astrophysics Source Code Library, record ascl:1010.083},
         year = 2010,
        month = oct,
          eid = {ascl:1010.083},
archivePrefix = {ascl},
       eprint = {1010.083},
       adsurl = {https://ui.adsabs.harvard.edu/abs/2010ascl.soft10083P}
}

@ARTICLE{paxton2013modules,
       author = {{Paxton}, Bill and {Cantiello}, Matteo and {Arras}, Phil and {Bildsten}, Lars and {Brown}, Edward F. and {Dotter}, Aaron and {Mankovich}, Christopher and {Montgomery}, M.~H. and {Stello}, Dennis and {Timmes}, F.~X. and {Townsend}, Richard},
        title = "{Modules for Experiments in Stellar Astrophysics (MESA): Planets, Oscillations, Rotation, and Massive Stars}",
      journal = {\apjs},
         year = 2013,
        month = sep,
       volume = {208},
       number = {1},
          eid = {4},
        pages = {4},
          doi = {10.1088/0067-0049/208/1/4},
archivePrefix = {arXiv},
       eprint = {1301.0319},
 primaryClass = {astro-ph.SR},
       adsurl = {https://ui.adsabs.harvard.edu/abs/2013ApJS..208....4P}
}

@ARTICLE{paxton2015modules,
       author = {{Paxton}, Bill and {Marchant}, Pablo and {Schwab}, Josiah and {Bauer}, Evan B. and {Bildsten}, Lars and {Cantiello}, Matteo and {Dessart}, Luc and {Farmer}, R. and {Hu}, H. and {Langer}, N. and {Townsend}, R.~H.~D. and {Townsley}, Dean M. and {Timmes}, F.~X.},
        title = "{Modules for Experiments in Stellar Astrophysics (MESA): Binaries, Pulsations, and Explosions}",
      journal = {\apjs},
         year = 2015,
        month = sep,
       volume = {220},
       number = {1},
          eid = {15},
        pages = {15},
          doi = {10.1088/0067-0049/220/1/15},
archivePrefix = {arXiv},
       eprint = {1506.03146},
 primaryClass = {astro-ph.SR},
       adsurl = {https://ui.adsabs.harvard.edu/abs/2015ApJS..220...15P}
}

@ARTICLE{paxton2017modules,
       author = {{Paxton}, Bill and {Schwab}, Josiah and {Bauer}, Evan B. and {Bildsten}, Lars and {Blinnikov}, Sergei and {Duffell}, Paul and {Farmer}, R. and {Goldberg}, Jared A. and {Marchant}, Pablo and {Sorokina}, Elena and {Thoul}, Anne and {Townsend}, Richard H.~D. and {Timmes}, F.~X.},
        title = "{Modules for Experiments in Stellar Astrophysics (MESA): Convective Boundaries, Element Diffusion, and Massive Star Explosions}",
      journal = {\apjs},
         year = 2018,
        month = feb,
       volume = {234},
       number = {2},
          eid = {34},
        pages = {34},
          doi = {10.3847/1538-4365/aaa5a8},
archivePrefix = {arXiv},
       eprint = {1710.08424},
 primaryClass = {astro-ph.SR},
       adsurl = {https://ui.adsabs.harvard.edu/abs/2018ApJS..234...34P}
}

@ARTICLE{paxton2019modules,
       author = {{Paxton}, Bill and {Smolec}, R. and {Schwab}, Josiah and {Gautschy}, A. and {Bildsten}, Lars and {Cantiello}, Matteo and {Dotter}, Aaron and {Farmer}, R. and {Goldberg}, Jared A. and {Jermyn}, Adam S. and {Kanbur}, S.~M. and {Marchant}, Pablo and {Thoul}, Anne and {Townsend}, Richard H.~D. and {Wolf}, William M. and {Zhang}, Michael and {Timmes}, F.~X.},
        title = "{Modules for Experiments in Stellar Astrophysics (MESA): Pulsating Variable Stars, Rotation, Convective Boundaries, and Energy Conservation}",
      journal = {\apjs},
         year = 2019,
        month = jul,
       volume = {243},
       number = {1},
          eid = {10},
        pages = {10},
          doi = {10.3847/1538-4365/ab2241},
archivePrefix = {arXiv},
       eprint = {1903.01426},
 primaryClass = {astro-ph.SR},
       adsurl = {https://ui.adsabs.harvard.edu/abs/2019ApJS..243...10P}
}

@ARTICLE{van1996accretion,
       author = {{van Paradijs}, J.},
        title = "{On the Accretion Instability in Soft X-Ray Transients}",
      journal = {\apjl},
         year = 1996,
        month = jun,
       volume = {464},
        pages = {L139},
          doi = {10.1086/310100},
       adsurl = {https://ui.adsabs.harvard.edu/abs/1996ApJ...464L.139V}
}

@ARTICLE{iglesias1996updated,
       author = {{Iglesias}, Carlos A. and {Rogers}, Forrest J.},
        title = "{Updated Opal Opacities}",
      journal = {\apj},
         year = 1996,
        month = jun,
       volume = {464},
        pages = {943},
          doi = {10.1086/177381},
       adsurl = {https://ui.adsabs.harvard.edu/abs/1996ApJ...464..943I}
}

@ARTICLE{Han2004MNRAS.350.1301H,
       author = {{Han}, Z. and {Podsiadlowski}, Ph.},
        title = "{The single-degenerate channel for the progenitors of Type Ia supernovae}",
      journal = {\mnras},
         year = 2004,
        month = jun,
       volume = {350},
       number = {4},
        pages = {1301-1309},
          doi = {10.1111/j.1365-2966.2004.07713.x},
archivePrefix = {arXiv},
       eprint = {astro-ph/0309618},
 primaryClass = {astro-ph},
       adsurl = {https://ui.adsabs.harvard.edu/abs/2004MNRAS.350.1301H}
}

@ARTICLE{GKPer2017,
       author = {{Zemko}, P. and {Orio}, M. and {Luna}, G.~J.~M. and {Mukai}, K. and {Evans}, P.~A. and {Bianchini}, A.},
        title = "{Multimission observations of the old nova GK Per during the 2015 outburst}",
      journal = {\mnras},
         year = 2017,
        month = jul,
       volume = {469},
       number = {1},
        pages = {476-491},
          doi = {10.1093/mnras/stx851},
archivePrefix = {arXiv},
       eprint = {1705.07707},
 primaryClass = {astro-ph.SR},
       adsurl = {https://ui.adsabs.harvard.edu/abs/2017MNRAS.469..476Z}
}

@ARTICLE{kato2021v606,
       author = {{Kato}, Taichi and {Kojiguchi}, Naoto},
        title = "{V606 Aql (Nova Aquilae 1899) is now a dwarf nova}",
      journal = {arXiv e-prints},
         year = 2021,
        month = jul,
          eid = {arXiv:2107.07055},
        pages = {arXiv:2107.07055},
          doi = {10.48550/arXiv.2107.07055},
archivePrefix = {arXiv},
       eprint = {2107.07055},
 primaryClass = {astro-ph.SR},
       adsurl = {https://ui.adsabs.harvard.edu/abs/2021arXiv210707055K}
}

@ARTICLE{murphy2022v392,
       author = {{Murphy-Glaysher}, F.~J. and {Darnley}, M.~J. and {Harvey}, {\'E}. J. and {Newsam}, A.~M. and {Page}, K.~L. and {Starrfield}, S. and {Wagner}, R.~M. and {Woodward}, C.~E. and {Terndrup}, D.~M. and {Kafka}, S. and {Arranz Heras}, T. and {Berardi}, P. and {Bertrand}, E. and {Biernikowicz}, R. and {Boussin}, C. and {Boyd}, D. and {Buchet}, Y. and {Bundas}, M. and {Coulter}, D. and {Dejean}, D. and {Diepvens}, A. and {Dvorak}, S. and {Edlin}, J. and {Eenmae}, T. and {Eggenstein}, H. and {Fournier}, R. and {Garde}, O. and {Gout}, J. and {Janzen}, D. and {Jordanov}, P. and {Kiiskinen}, H. and {Lane}, D. and {Larochelle}, R. and {Leadbeater}, R. and {Mankel}, D. and {Martineau}, G. and {Miller}, I. and {Modic}, R. and {Montier}, J. and {Morales Aimar}, M. and {Muyllaert}, E. and {Naves Nogues}, R. and {O'Keeffe}, D. and {Oksanen}, A. and {Pyatnytskyy}, M. and {Rast}, R. and {Rodgers}, B. and {Rodriguez Perez}, D. and {Schorr}, F. and {Schwendeman}, E. and {Shadick}, S. and {Sharpe}, S. and {Sold{\'a}n Alfaro}, F. and {Sove}, T. and {Stone}, G. and {Tordai}, T. and {Venne}, R. and {Vollmann}, W. and {Vrastak}, M. and {Wenzel}, K.},
        title = "{V392 Persei: A {\ensuremath{\gamma}}-ray bright nova eruption from a known dwarf nova}",
      journal = {\mnras},
         year = 2022,
        month = aug,
       volume = {514},
       number = {4},
        pages = {6183-6202},
          doi = {10.1093/mnras/stac1577},
archivePrefix = {arXiv},
       eprint = {2206.03443},
 primaryClass = {astro-ph.HE},
       adsurl = {https://ui.adsabs.harvard.edu/abs/2022MNRAS.514.6183M}
}

@ARTICLE{schwab2016evolution,
       author = {{Schwab}, Josiah and {Quataert}, Eliot and {Kasen}, Daniel},
        title = "{The evolution and fate of super-Chandrasekhar mass white dwarf merger remnants}",
      journal = {\mnras},
         year = 2016,
        month = dec,
       volume = {463},
       number = {4},
        pages = {3461-3475},
          doi = {10.1093/mnras/stw2249},
archivePrefix = {arXiv},
       eprint = {1606.02300},
 primaryClass = {astro-ph.SR},
       adsurl = {https://ui.adsabs.harvard.edu/abs/2016MNRAS.463.3461S}
}

@ARTICLE{wu2018accreting,
       author = {{Wu}, Cheng-Yuan and {Wang}, Bo},
        title = "{Accreting CO material onto ONe white dwarfs towards accretion-induced collapse}",
      journal = {Research in Astronomy and Astrophysics},
         year = 2018,
        month = mar,
       volume = {18},
       number = {3},
          eid = {036},
        pages = {036},
          doi = {10.1088/1674-4527/18/3/36},
archivePrefix = {arXiv},
       eprint = {1710.04417},
 primaryClass = {astro-ph.SR},
       adsurl = {https://ui.adsabs.harvard.edu/abs/2018RAA....18...36W}
}

@ARTICLE{wang2018single,
       author = {{Wang}, Bo},
        title = "{The single-degenerate model for the progenitors of accretion-induced collapse events}",
      journal = {\mnras},
         year = 2018,
        month = nov,
       volume = {481},
       number = {1},
        pages = {439-446},
          doi = {10.1093/mnras/sty2278},
archivePrefix = {arXiv},
       eprint = {1808.05992},
 primaryClass = {astro-ph.SR},
       adsurl = {https://ui.adsabs.harvard.edu/abs/2018MNRAS.481..439W}
}

@ARTICLE{wu2019off,
       author = {{Wu}, Chengyuan and {Wang}, Bo},
        title = "{Off-centre carbon burning in He-accreting carbon-oxygen white dwarfs}",
      journal = {\mnras},
         year = 2019,
        month = jul,
       volume = {486},
       number = {3},
        pages = {2977-2981},
          doi = {10.1093/mnras/stz1028},
archivePrefix = {arXiv},
       eprint = {1904.05130},
 primaryClass = {astro-ph.SR},
       adsurl = {https://ui.adsabs.harvard.edu/abs/2019MNRAS.486.2977W}
}

@ARTICLE{liu2020formation,
       author = {{Liu}, D. and {Wang}, B.},
        title = "{The formation of single neutron stars from double white-dwarf mergers via accretion-induced collapse}",
      journal = {\mnras},
         year = 2020,
        month = may,
       volume = {494},
       number = {3},
        pages = {3422-3431},
          doi = {10.1093/mnras/staa963},
archivePrefix = {arXiv},
       eprint = {2004.03157},
 primaryClass = {astro-ph.SR},
       adsurl = {https://ui.adsabs.harvard.edu/abs/2020MNRAS.494.3422L}
}

@ARTICLE{godon2006hubble,
       author = {{Godon}, Patrick and {Sion}, Edward M. and {Cheng}, Fuhua and {Long}, Knox S. and {G{\"a}nsicke}, Boris T. and {Szkody}, Paula},
        title = "{Hubble Space Telescope STIS Spectroscopy and Modeling of the Long-Term Cooling of WZ Sagittae following the 2001 July Outburst}",
      journal = {\apj},
         year = 2006,
        month = may,
       volume = {642},
       number = {2},
        pages = {1018-1028},
          doi = {10.1086/501039},
archivePrefix = {arXiv},
       eprint = {astro-ph/0602167},
 primaryClass = {astro-ph},
       adsurl = {https://ui.adsabs.harvard.edu/abs/2006ApJ...642.1018G}
}

@ARTICLE{shara2007ancient,
       author = {{Shara}, Michael M. and {Martin}, Christopher D. and {Seibert}, Mark and {Rich}, R. Michael and {Salim}, Samir and {Reitzel}, David and {Schiminovich}, David and {Deliyannis}, Constantine P. and {Sarrazine}, Angela R. and {Kulkarni}, Shri R. and {Ofek}, Eran O. and {Brosch}, Noah and {L{\'e}pine}, Sebastien and {Zurek}, David and {De Marco}, Orsola and {Jacoby}, George},
        title = "{An ancient nova shell around the dwarf nova Z Camelopardalis}",
      journal = {Nat},
         year = 2007,
        month = mar,
       volume = {446},
       number = {7132},
        pages = {159-162},
          doi = {10.1038/nature05576},
       adsurl = {https://ui.adsabs.harvard.edu/abs/2007Natur.446..159S}
}

@ARTICLE{shara2012cnc,
       author = {{Shara}, Michael M. and {Mizusawa}, Trisha and {Wehinger}, Peter and {Zurek}, David and {Martin}, Christopher D. and {Neill}, James D. and {Forster}, Karl and {Seibert}, Mark},
        title = "{AT Cnc: A Second Dwarf Nova with a Classical Nova Shell}",
      journal = {\apj},
         year = 2012,
        month = oct,
       volume = {758},
       number = {2},
          eid = {121},
        pages = {121},
          doi = {10.1088/0004-637X/758/2/121},
archivePrefix = {arXiv},
       eprint = {1208.1280},
 primaryClass = {astro-ph.SR},
       adsurl = {https://ui.adsabs.harvard.edu/abs/2012ApJ...758..121S}
}

@ARTICLE{joung2009dependence,
       author = {{Joung}, M. Ryan and {Mac Low}, Mordecai-Mark and {Bryan}, Greg L.},
        title = "{Dependence of Interstellar Turbulent Pressure on Supernova Rate}",
      journal = {\apj},
         year = 2009,
        month = oct,
       volume = {704},
       number = {1},
        pages = {137-149},
          doi = {10.1088/0004-637X/704/1/137},
archivePrefix = {arXiv},
       eprint = {0811.3747},
 primaryClass = {astro-ph},
       adsurl = {https://ui.adsabs.harvard.edu/abs/2009ApJ...704..137J}
}

@ARTICLE{strickland2009supernova,
       author = {{Strickland}, David K. and {Heckman}, Timothy M.},
        title = "{Supernova Feedback Efficiency and Mass Loading in the Starburst and Galactic Superwind Exemplar M82}",
      journal = {\apj},
         year = 2009,
        month = jun,
       volume = {697},
       number = {2},
        pages = {2030-2056},
          doi = {10.1088/0004-637X/697/2/2030},
archivePrefix = {arXiv},
       eprint = {0903.4175},
 primaryClass = {astro-ph.CO},
       adsurl = {https://ui.adsabs.harvard.edu/abs/2009ApJ...697.2030S}
}

@ARTICLE{wang2015super,
       author = {{Wang}, B. and {Ma}, X. and {Liu}, D.-D. and {Liu}, Z.-W. and {Wu}, C.-Y. and {Zhang}, J.-J. and {Han}, Z.},
        title = "{Super-Eddington wind scenario for the progenitors of type Ia supernovae: binary population synthesis calculations}",
      journal = {\aap},
         year = 2015,
        month = apr,
       volume = {576},
          eid = {A86},
        pages = {A86},
          doi = {10.1051/0004-6361/201425294},
archivePrefix = {arXiv},
       eprint = {1502.06307},
 primaryClass = {astro-ph.SR},
       adsurl = {https://ui.adsabs.harvard.edu/abs/2015A&A...576A..86W}
}

@ARTICLE{brooks2016carbon,
       author = {{Brooks}, Jared and {Bildsten}, Lars and {Schwab}, Josiah and {Paxton}, Bill},
        title = "{Carbon Shell or Core Ignitions in White Dwarfs Accreting from Helium Stars}",
      journal = {\apj},
         year = 2016,
        month = apr,
       volume = {821},
       number = {1},
          eid = {28},
        pages = {28},
          doi = {10.3847/0004-637X/821/1/28},
archivePrefix = {arXiv},
       eprint = {1602.05586},
 primaryClass = {astro-ph.HE},
       adsurl = {https://ui.adsabs.harvard.edu/abs/2016ApJ...821...28B}
}

@ARTICLE{wu2020formation,
       author = {{Wu}, Chengyuan and {Wang}, Bo and {Wang}, Xiaofeng and {Maeda}, Keiichi and {Mazzali}, Paolo},
        title = "{The formation of type Ia supernovae from carbon-oxygen-silicon white dwarfs}",
      journal = {\mnras},
         year = 2020,
        month = jun,
       volume = {495},
       number = {1},
        pages = {1445-1460},
          doi = {10.1093/mnras/staa1277},
archivePrefix = {arXiv},
       eprint = {2005.01967},
 primaryClass = {astro-ph.HE},
       adsurl = {https://ui.adsabs.harvard.edu/abs/2020MNRAS.495.1445W}
}

@ARTICLE{wong2019evolution,
       author = {{Wong}, Tin Long Sunny and {Schwab}, Josiah},
        title = "{Evolution of Helium Star-White Dwarf Binaries Leading up to Thermonuclear Supernovae}",
      journal = {\apj},
         year = 2019,
        month = jun,
       volume = {878},
       number = {2},
          eid = {100},
        pages = {100},
          doi = {10.3847/1538-4357/ab1b49},
archivePrefix = {arXiv},
       eprint = {1901.04512},
 primaryClass = {astro-ph.SR},
       adsurl = {https://ui.adsabs.harvard.edu/abs/2019ApJ...878..100W}
}

@ARTICLE{schwab2021evolutionary,
       author = {{Schwab}, Josiah},
        title = "{Evolutionary Models for the Remnant of the Merger of Two Carbon-Oxygen Core White Dwarfs}",
      journal = {\apj},
         year = 2021,
        month = jan,
       volume = {906},
       number = {1},
          eid = {53},
        pages = {53},
          doi = {10.3847/1538-4357/abc87e},
archivePrefix = {arXiv},
       eprint = {2011.03546},
 primaryClass = {astro-ph.SR},
       adsurl = {https://ui.adsabs.harvard.edu/abs/2021ApJ...906...53S}
}

@ARTICLE{whelan1973binaries,
       author = {{Whelan}, John and {Iben}, Jr., Icko},
        title = "{Binaries and Supernovae of Type I}",
      journal = {\apj},
         year = 1973,
        month = dec,
       volume = {186},
        pages = {1007-1014},
          doi = {10.1086/152565},
       adsurl = {https://ui.adsabs.harvard.edu/abs/1973ApJ...186.1007W}
}

@ARTICLE{maoz2014observational,
       author = {{Maoz}, Dan and {Mannucci}, Filippo and {Nelemans}, Gijs},
        title = "{Observational Clues to the Progenitors of Type Ia Supernovae}",
      journal = {\araa},
         year = 2014,
        month = aug,
       volume = {52},
        pages = {107-170},
          doi = {10.1146/annurev-astro-082812-141031},
archivePrefix = {arXiv},
       eprint = {1312.0628},
 primaryClass = {astro-ph.CO},
       adsurl = {https://ui.adsabs.harvard.edu/abs/2014ARA&A..52..107M}
}

@ARTICLE{maoz2012type,
       author = {{Maoz}, D. and {Mannucci}, F.},
        title = "{Type-Ia Supernova Rates and the Progenitor Problem: A Review}",
      journal = {\pasa},
         year = 2012,
        month = jan,
       volume = {29},
       number = {4},
        pages = {447-465},
          doi = {10.1071/AS11052},
archivePrefix = {arXiv},
       eprint = {1111.4492},
 primaryClass = {astro-ph.CO},
       adsurl = {https://ui.adsabs.harvard.edu/abs/2012PASA...29..447M}
}

@ARTICLE{maoz2017star,
       author = {{Maoz}, Dan and {Graur}, Or},
        title = "{Star Formation, Supernovae, Iron, and {\ensuremath{\alpha}}: Consistent Cosmic and Galactic Histories}",
      journal = {\apj},
         year = 2017,
        month = oct,
       volume = {848},
       number = {1},
          eid = {25},
        pages = {25},
          doi = {10.3847/1538-4357/aa8b6e},
archivePrefix = {arXiv},
       eprint = {1703.04540},
 primaryClass = {astro-ph.HE},
       adsurl = {https://ui.adsabs.harvard.edu/abs/2017ApJ...848...25M}
}

@ARTICLE{leung2018explosive,
       author = {{Leung}, Shing-Chi and {Nomoto}, Ken'ichi},
        title = "{Explosive Nucleosynthesis in Near-Chandrasekhar-mass White Dwarf Models for Type Ia Supernovae: Dependence on Model Parameters}",
      journal = {\apj},
         year = 2018,
        month = jul,
       volume = {861},
       number = {2},
          eid = {143},
        pages = {143},
          doi = {10.3847/1538-4357/aac2df},
archivePrefix = {arXiv},
       eprint = {1710.04254},
 primaryClass = {astro-ph.SR},
       adsurl = {https://ui.adsabs.harvard.edu/abs/2018ApJ...861..143L}
}

@ARTICLE{Pakmor2010,
       author = {{Pakmor}, R{\"u}diger and {Kromer}, Markus and {R{\"o}pke}, Friedrich K. and {Sim}, Stuart A. and {Ruiter}, Ashley J. and {Hillebrandt}, Wolfgang},
        title = "{Sub-luminous type Ia supernovae from the mergers of equal-mass white dwarfs with mass \raisebox{-0.5ex}\textasciitilde0.9M$_{solar}$}",
      journal = {Nat},
         year = 2010,
        month = jan,
       volume = {463},
       number = {7277},
        pages = {61-64},
          doi = {10.1038/nature08642},
archivePrefix = {arXiv},
       eprint = {0911.0926},
 primaryClass = {astro-ph.HE},
       adsurl = {https://ui.adsabs.harvard.edu/abs/2010Natur.463...61P}
}

@ARTICLE{pakmor2011,
       author = {{Pakmor}, R. and {Hachinger}, S. and {R{\"o}pke}, F.~K. and {Hillebrandt}, W.},
        title = "{Violent mergers of nearly equal-mass white dwarf as progenitors of subluminous Type Ia supernovae}",
      journal = {\aap},
         year = 2011,
        month = apr,
       volume = {528},
          eid = {A117},
        pages = {A117},
          doi = {10.1051/0004-6361/201015653},
archivePrefix = {arXiv},
       eprint = {1102.1354},
 primaryClass = {astro-ph.SR},
       adsurl = {https://ui.adsabs.harvard.edu/abs/2011A&A...528A.117P}
}

@ARTICLE{pakmor2012,
       author = {{Pakmor}, R. and {Kromer}, M. and {Taubenberger}, S. and {Sim}, S.~A. and {R{\"o}pke}, F.~K. and {Hillebrandt}, W.},
        title = "{Normal Type Ia Supernovae from Violent Mergers of White Dwarf Binaries}",
      journal = {\apjl},
         year = 2012,
        month = mar,
       volume = {747},
       number = {1},
          eid = {L10},
        pages = {L10},
          doi = {10.1088/2041-8205/747/1/L10},
archivePrefix = {arXiv},
       eprint = {1201.5123},
 primaryClass = {astro-ph.HE},
       adsurl = {https://ui.adsabs.harvard.edu/abs/2012ApJ...747L..10P}
}

@ARTICLE{yoon2007,
       author = {{Yoon}, S.-C. and {Podsiadlowski}, Ph. and {Rosswog}, S.},
        title = "{Remnant evolution after a carbon-oxygen white dwarf merger}",
      journal = {\mnras},
         year = 2007,
        month = sep,
       volume = {380},
       number = {3},
        pages = {933-948},
          doi = {10.1111/j.1365-2966.2007.12161.x},
archivePrefix = {arXiv},
       eprint = {0704.0297},
 primaryClass = {astro-ph},
       adsurl = {https://ui.adsabs.harvard.edu/abs/2007MNRAS.380..933Y}
}

@ARTICLE{fink2007,
       author = {{Fink}, M. and {Hillebrandt}, W. and {R{\"o}pke}, F.~K.},
        title = "{Double-detonation supernovae of sub-Chandrasekhar mass white dwarfs}",
      journal = {\aap},
         year = 2007,
        month = dec,
       volume = {476},
       number = {3},
        pages = {1133-1143},
          doi = {10.1051/0004-6361:20078438},
archivePrefix = {arXiv},
       eprint = {0710.5486},
 primaryClass = {astro-ph},
       adsurl = {https://ui.adsabs.harvard.edu/abs/2007A&A...476.1133F}
}

@ARTICLE{fink2010,
       author = {{Fink}, M. and {R{\"o}pke}, F.~K. and {Hillebrandt}, W. and {Seitenzahl}, I.~R. and {Sim}, S.~A. and {Kromer}, M.},
        title = "{Double-detonation sub-Chandrasekhar supernovae: can minimum helium shell masses detonate the core?}",
      journal = {\aap},
         year = 2010,
        month = may,
       volume = {514},
          eid = {A53},
        pages = {A53},
          doi = {10.1051/0004-6361/200913892},
archivePrefix = {arXiv},
       eprint = {1002.2173},
 primaryClass = {astro-ph.SR},
       adsurl = {https://ui.adsabs.harvard.edu/abs/2010A&A...514A..53F}
}

@ARTICLE{boos2021,
       author = {{Boos}, Samuel J. and {Townsley}, Dean M. and {Shen}, Ken J. and {Caldwell}, Spencer and {Miles}, Broxton J.},
        title = "{Multidimensional Parameter Study of Double Detonation Type Ia Supernovae Originating from Thin Helium Shell White Dwarfs}",
      journal = {\apj},
         year = 2021,
        month = oct,
       volume = {919},
       number = {2},
          eid = {126},
        pages = {126},
          doi = {10.3847/1538-4357/ac07a2},
archivePrefix = {arXiv},
       eprint = {2101.12330},
 primaryClass = {astro-ph.HE},
       adsurl = {https://ui.adsabs.harvard.edu/abs/2021ApJ...919..126B}
}

@ARTICLE{townsley2004,
       author = {{Townsley}, Dean M. and {Bildsten}, Lars},
        title = "{Theoretical Modeling of the Thermal State of Accreting White Dwarfs Undergoing Classical Nova Cycles}",
      journal = {\apj},
         year = 2004,
        month = jan,
       volume = {600},
       number = {1},
        pages = {390-403},
          doi = {10.1086/379647},
archivePrefix = {arXiv},
       eprint = {astro-ph/0306080},
 primaryClass = {astro-ph},
       adsurl = {https://ui.adsabs.harvard.edu/abs/2004ApJ...600..390T}
}

@ARTICLE{piro2005,
       author = {{Piro}, Anthony L. and {Arras}, Phil and {Bildsten}, Lars},
        title = "{White Dwarf Heating and Subsequent Cooling in Dwarf Nova Outbursts}",
      journal = {\apj},
         year = 2005,
        month = jul,
       volume = {628},
       number = {1},
        pages = {401-410},
          doi = {10.1086/430588},
archivePrefix = {arXiv},
       eprint = {astro-ph/0503641},
 primaryClass = {astro-ph},
       adsurl = {https://ui.adsabs.harvard.edu/abs/2005ApJ...628..401P}
}

@ARTICLE{townsley2009,
       author = {{Townsley}, Dean M. and {G{\"a}nsicke}, Boris T.},
        title = "{Cataclysmic Variable Primary Effective Temperatures: Constraints on Binary Angular Momentum Loss}",
      journal = {\apj},
         year = 2009,
        month = mar,
       volume = {693},
       number = {1},
        pages = {1007-1021},
          doi = {10.1088/0004-637X/693/1/1007},
archivePrefix = {arXiv},
       eprint = {0811.2447},
 primaryClass = {astro-ph},
       adsurl = {https://ui.adsabs.harvard.edu/abs/2009ApJ...693.1007T}
}

@ARTICLE{yaron2005,
       author = {{Yaron}, O. and {Prialnik}, D. and {Shara}, M.~M. and {Kovetz}, A.},
        title = "{An Extended Grid of Nova Models. II. The Parameter Space of Nova Outbursts}",
      journal = {\apj},
         year = 2005,
        month = apr,
       volume = {623},
       number = {1},
        pages = {398-410},
          doi = {10.1086/428435},
archivePrefix = {arXiv},
       eprint = {astro-ph/0503143},
 primaryClass = {astro-ph},
       adsurl = {https://ui.adsabs.harvard.edu/abs/2005ApJ...623..398Y}
}

@ARTICLE{yoon2004,
       author = {{Yoon}, S.-C. and {Langer}, N.},
        title = "{Presupernova evolution of accreting white dwarfs with rotation}",
      journal = {\aap},
         year = 2004,
        month = may,
       volume = {419},
        pages = {623-644},
          doi = {10.1051/0004-6361:20035822},
archivePrefix = {arXiv},
       eprint = {astro-ph/0402287},
 primaryClass = {astro-ph},
       adsurl = {https://ui.adsabs.harvard.edu/abs/2004A&A...419..623Y}
}
\bibliographystyle{aa}
\end{document}